\documentclass{aa}  

\usepackage{graphicx}
\usepackage{txfonts}
\usepackage{lipsum}
\usepackage{subcaption}         
\usepackage{lscape}            
\usepackage{placeins}          
\usepackage{xcolor}

\usepackage{comment}
\usepackage{hyperref}
\usepackage{multirow}
\newcommand{\orcidlink}[1]{\protect\href{https://orcid.org/#1}{\protect\includegraphics[width=8pt]{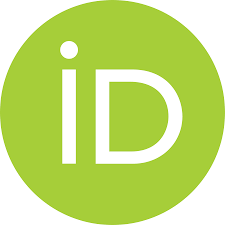}}}

\begin{document}

   \title{Evolving stellar wind environment around M dwarfs and its influence on magnetic star-planet interactions}

\titlerunning{Evolving stellar wind environment around M dwarfs}

   \author{S. W. Petyniak\inst{1}\fnmsep\thanks{Corresponding author \email{petyniak@strw.leidenuniv.nl}}\orcidlink{0009-0006-7449-6055}
        \and S. Bellotti \inst{1, 2} \orcidlink{0000-0002-2558-6920}
        \and A. A. Vidotto \inst{1} \orcidlink{0000-0001-5371-2675}
        \and P. I. Cristofari \inst{1, 3} \orcidlink{0000-0003-4019-0630}
        \and J. F. Donati \inst{2}\orcidlink{0000-0001-5541-2887}
        \and A. Biswas \inst{4}\orcidlink{0000-0002-1741-6286}
        \and C.~P.~Folsom \inst{5}\orcidlink{0000-0002-9023-7890}
        }
    \authorrunning{Petyniak et al.}
   \institute{Leiden Observatory, Leiden University, PO Box 9513, 2300 RA Leiden, the Netherlands
   \and Institut de Recherche en Astrophysique et Planétologie, Université de Toulouse, CNRS, IRAP/UMR 5277, 14 Avenue Edouard Belin, 31400 Toulouse, France
   \and Center for Astrophysics | Harvard \& Smithsonian, 60 Garden St., Cambridge, MA 02138, USA
   \and National Centre for Radio Astrophysics, Tata Institute of Fundamental Research, Ganeshkhind, Pune-411007, India
   \and Tartu Observatory, University of Tartu, Observatooriumi 1, T\~oravere 61602, Estonia
   }

   \date{\swpt{Received 11 June 2026 / Accepted 5 November 2026}}

\newcommand{\aav}[1]{{\color{black}#1}}
\newcommand{\swp}[1]{{\color{black}#1}}
\newcommand{\swpt}[1]{{\color{black}#1}}

\newcommand{\valrange}[2]{
  \begin{tabular}[c]{@{}c@{}}
    #1\\[-2pt]
    {\scriptsize #2}
  \end{tabular}
}

\newcommand{\planetcell}[2]{
  \multirow{#1}{*}{
    \centering
    \vphantom{#2}
    #2
  }
}
 
  \abstract
   {M dwarfs are the most abundant group of stars in the Galaxy, and they are expected to host a large number of exoplanets orbiting in the habitable zone. It is believed that stellar winds of M dwarfs can have a substantial \swp{effect} on these exoplanets and \swp{on the} observations of these systems. However, a systematic investigation of how the stellar winds of M dwarfs varies on timescales of \swp{a} few months to years and how this impacts exoplanets around these stars is missing.}
   {We aim to investigate how  variability of magnetic field properties of M dwarfs affects their stellar winds, \aav{which} is essential for future studies of star-planet interactions (SPIs) and planetary habitability. In particular, \aav{such} variability might shed light on whether SPI signatures are going to repeat.} 
   {We have \swp{compiled} a total of 18 magnetic field maps \swp{for} six M dwarfs, with masses between 0.1 and 0.6 $M_\odot$, and rotation periods between 2.7 and 175.6 days. We use these maps as input of 3D \swp{magnetohydrodynamic} stellar wind simulations. We derive the mass-loss rate, the angular momentum-loss rate, and the Alfvén radius. For the stars with confirmed exoplanets (\swp{DS Leo}, YZ Cet, and GJ 1151) we investigate the variability of stellar wind properties at the orbit.}
   {We find mass-loss rates between \swpt{$1.0\times10^{-14}~M_\odot~{\rm yr^{-1}}$ and $2.3 \times 10^{-13}~M_\odot~{\rm yr^{-1}}$} and angular momentum-loss rates between $4.6 \times 10^{27}~{\rm erg}$ and $4.9 \times 10^{31}~{\rm erg}$. For individual stars the \swpt{Alfvén radius, mass-loss rate, and angular momentum-loss rate varied with time by up to a factor of 1.5, 3.8, and 7.4 respectively}.  Our results follow the observational relation between mass-loss rate and X-ray flux. Due to the time-varying Alfvén surface, exoplanets orbiting \swp{DS Leo} and YZ Cet move in and out of \aav{the} sub-Alfvénic regime, affecting the nature of star-planet interactions and prospects for their detection.
   }
   {}


   \maketitle
\nolinenumbers

\section{Introduction}
M dwarfs constitute the most common group of stars in the Galaxy, with \aav{at least 70\%} of stars on the main sequence being of that spectral type \aav{\citep[e.g.,][]{2004ASPC..318..159H}}. M dwarfs host many exoplanets with estimates of about \swpt{1--1.5} planet per star, most of which are less than 10 Earth masses \aav{\citep[e.g.,][see also \citealt{Bonfils2013} for an earlier study]{Sabotta2021, 2025A&A...696A.101K}.} \aav{Such an} abundance makes M dwarfs \aav{attractive} targets for missions characterizing exoplanets such as JWST \citep{Gardner2006, Morley2017} or the upcoming \swp{\it Ariel} mission \citep{Tinetti2018, Edwards2019}. Furthermore, many \aav{M-dwarf planets}  orbit in the habitable zone \citep[][]{Bonfils2013, Dressing2015}.

Since the effective temperature of M dwarfs is lower than that of the Sun, their habitable zones are closer to the star \swpt{\citep[e.g.,][]{1993Icar..101..108K, Kopparapu2013b, 2014ApJ...787L..29K}}. \aav{Such a} proximity leads to debates whether exoplanets orbiting M dwarfs can have favorable conditions for life \aav{\citep[e.g.,][]{2007AsBio...7...30T,Scalo2007,Shields2016}}. M dwarfs remain magnetically active for a large part of their lives \aav{\citep[e.g.,][]{2008AJ....135..785W,2009ApJ...692..538R,2011MNRAS.413.2595S,Wright2018}}, which can have adverse effects on life \citep[][]{Atri2017,Tiley2019} and cause atmospheric loss in exoplanets \aav{\citep[e.g.,][]{2010A&A...511L...8S, 2016MNRAS.459.4088O, Kay2016, Yamashiki2019, Airapetian2019,Carolan2020}}. Studies have shown that exoplanets around M dwarfs may be exposed to quiescent stellar wind conditions that are stronger than what we observe with Solar System planets \citep[][]{Vidotto2013, Garraffo2016}. The magnetic activity of a \aav{low-mass star} is linked to its magnetic field, which ultimately affects the stellar wind and the surrounding space environment around it \citep[][]{See2016,Vidotto2021}. 

For  stars, magnetic field can be observed using the Zeeman effect \citep[][]{Zeeman1897}. Zeeman Doppler imaging \citep[ZDI; ][]{Semel1989,DonatiBrown1997, Donati1997_LSD} applied to circular and linear polarization spectra can reveal the topology of the large-scale magnetic field. 
 Spectropolarimetric observations have shown that magnetic fields of M dwarfs organized on both small and large spatial scales  undergo temporal variations, potentially reminiscent of magnetic cycles \citep[e.g.,][]{Bellotti2023, Bellotti2024b, Cristofari2025}. Other observations, including activity indicators and photometry, have shown that M dwarfs undergo activity cycles on timescales of a few years \aav{\citep[e.g.,][]{SuarezMascareno2016, 2019A&A...621A.126D, Mignon2023, Oviedo2026}}.

A key consideration to understand the relation between the star and the planet is the position of an exoplanet relative to the Alfvén surface, which is the boundary where plasma achieves Alfvén speed \swpt{--} the speed at which Alfvén waves propagate. Stellar wind accelerates as it moves away from the star \citep[][]{Parker1958}, eventually reaching speeds exceeding Alfvén speed. Planets orbiting inside of the Alfvén surface are referred to as being in the sub-Alfvénic regime, while those outside are super-Alfvénic.

Depending on the planetary orbit and stellar wind conditions, different star-planet interactions (SPIs) can occur that are not observed in the Solar System \citep[][]{Vidotto2025}. For example, planets in the sub-Alfvénic regime can magnetically connect to the star, potentially \swp{influencing chromospheric emission by producing hot spots \citep{Shkolnik2003, Shkolnik2005, Shkolnik2008, Cauley2018, ZapateroOsorio2026, Revilla2026}, as well as} affecting flaring phenomena \citep[][]{Lanza2018, Loyd2023, Ilin2025}, planetary atmospheric escape \citep[][]{Lanza2013, Presa2024, Presa2026}. On \swp{the} other hand, magnetized planets orbiting in the super-Alfvénic regime can form bow shocks, which divert the stellar wind shielding the surface from energetic particles \citep[][]{KivelsonRussel}. Such protection is expected to be an important factor for planetary habitability \citep[][]{Khodachenko2007, Vidotto2013}. However, questions remain as to whether planetary magnetic fields enhance or lower atmospheric escape \citep[see e.g.][]{Blackman2018, Gunell2018}. Understanding the evolution of the stellar magnetic field, stellar wind conditions and Alfvén surface is essential both for detection and interpretation of SPIs \citep[e.g.,][]{Presa2026}.

In this work, we characterize the variability of the stellar environment around M dwarfs following \cite{Smith2026}. While previous analyses used large-scale topology reconstructed for one epoch per star \citep[e.g.][]{Vidotto2014b, Chebly2023}, we characterized the temporal variations of the stellar wind by using multiple epochs for each M dwarf in the sample. For stars hosting confirmed exoplanets, we analyzed the relation between their orbit and the Alfvén surface, identifying different regimes of SPIs. We also sampled stellar wind density, velocity, magnetic field strength, and temperature at the orbits of these exoplanets.

The paper is structured as follows. In Sect. \ref{section:our_sample}, we describe the sample of stars investigated in this work, and briefly explain the ZDI maps used for the simulations. \swpt{In Sect. \ref{section:stellar_wind} we describe the ZDI-driven stellar wind models and analyze the simulation outputs for individual epochs, the evolution of stellar wind properties, and the environment around known planets.} In Sect. \ref{section:discussion} we discuss our findings. In Sect. \ref{section:conclusions} we summarize our results and draw our conclusions.

\section{Sample}\label{section:our_sample}

\subsection{Sample of stars}
In this work we used published magnetic field maps of M dwarfs listed in Table \ref{tab:sample_properties}. The maps were obtained using ZDI \citep{Semel1989,DonatiBrown1997} applied to near-infrared spectropolarimetric observations conducted with the SpectroPolarimètre InfraRouge \citep[SPIRou;][]{Donati2020}. For more details on the magnetic field reconstruction, see Table \ref{tab:sample_properties} and references therein.

Theoretical models indicate that the boundary between fully and partially convective stars occurs at approximately \swp{$0.35\,M_\odot$} \citep{Dorman1989, Chabrier1997, Rabus2019}. With this range, CN Leo, YZ Cet, \swp{GJ} 905, and GJ 1151 are expected to be fully convective; \swp{DS Leo} partially convective; and EV Lac lies near the transition between the two regimes. 
 
Fully convective stars do not posses the tachocline \swp{--} the thin region separating the radiative and convective layers of a star. In the Sun, this is expected to be a key component of the magnetic cycle \citep[][]{Charbonneau2010}. 

\swp{DS Leo}, YZ Cet, and GJ 1151 are known to host exoplanets, orbital properties of which are summarized in Table \ref{tab:exoplanet_orbits}. \swpt{The semi-major axis is reported in both astronomical units and the host star radii units. In case the referenced work fixes eccentricity instead of fitting we indicate that instead of providing an uncertainty. For the YZ Cet system, the parameters reported by \citet{Stock2020} are sufficiently close to those reported by \cite{Astudillo-Defru2017} that the choice between the two does not meaningfully affect the results presented in this work.} We analyzed stellar wind environment around exoplanets \swpt{orbitting DS Leo, YZ Cet, and GJ 1151} in \swp{Sect.} \ref{section:exoplanet_sw}. We do not consider candidate exoplanets in this work\swp{, which can be found on NASA Exoplanet Archive}\footnote{\swp{\url{ https://exoplanetarchive.ipac.caltech.edu/}}}. Recent radio detections from YZ Cet show that this system is a candidate for sub-Alfvénic star-planet interactions \citep[][]{Pineda2023, Trigilio2023, Pineda2026}. Characterizing its stellar wind environment is essential to provide context for these detections and aid future modeling efforts.

\begin{table*}[th!]
\caption{Fundamental properties of stars investigated in this paper, along with number of available magnetic field maps.}
\centering
\begin{tabular}{llccccc}
\hline\hline Star &
  \begin{tabular}[c]{@{}l@{}}Spectral \\ \swpt{type}\end{tabular} & 
  \begin{tabular}[c]{@{}l@{}}\swp{$P_{\rm rot}$ {[}d{]}} \end{tabular} &
  \begin{tabular}[c]{@{}l@{}}\swpt{$R$} $[R_\odot]$\end{tabular} &
  \begin{tabular}[c]{@{}l@{}}\swpt{$M$} $[M_\odot]$\end{tabular} & \begin{tabular}[c]{@{}l@{}}\swp{$\log(L_{\rm X})$}\\ \swp{$[\log({\rm erg~s^{-1}})]$}
  \end{tabular} &
  \begin{tabular}[c]{@{}l@{}}Number of \\ ZDI \swp{maps}\end{tabular} \\ \hline
CN Leo & M5.5\swp{\,V} & 2.70\swp{$\pm0.01$}\textsuperscript{a} & 0.12\swp{$\pm0.02$}\textsuperscript{b}  & 0.10\swp{$\pm0.02$}\textsuperscript{b} & 26.97\textsuperscript{c} &4\textsuperscript{a}  \\
EV Lac & M3.5\swp{\,V} & 4.36\swp{$\pm0.01$}\textsuperscript{a}  & 0.30\swp{$\pm0.02$}\textsuperscript{b} & 0.32\swp{$\pm0.02$}\textsuperscript{b} & 28.99\textsuperscript{c} & 4\textsuperscript{a,d}
\\
\swp{DS Leo} & M1.0\swp{\,V} & 13.91\swp{$\pm0.01$}\textsuperscript{a} & 0.53\swp{$\pm0.02$}\textsuperscript{b}  & 0.58\swp{$\pm0.02$}\textsuperscript{b} & 28.37\textsuperscript{c} & 2\textsuperscript{a} \\
YZ Cet & M4.5\swp{\,V} & 68.40\swp{$\pm0.05$}\textsuperscript{e}  & \swp{$0.157\pm0.005$}\textsuperscript{e} & \swp{$0.142 \pm 0.010$}\textsuperscript{f} & 27.40\textsuperscript{c} & 2\textsuperscript{g,h} \\
\swp{GJ} 905 & M5.0\swp{\,V} & 114.30\swp{$\pm2.80$}\textsuperscript{i} & 0.17\swp{$\pm0.01$}\textsuperscript{j} & 0.15\swp{$\pm0.02$}\textsuperscript{j} & 27.10\textsuperscript{c} & 3\textsuperscript{k} \\
GJ 1151 & M4.5\swp{\,V} & 175.60\swp{$\pm4.90$}\textsuperscript{i} & 0.19\swp{$\pm0.01$}\textsuperscript{j} & 0.17\swp{$\pm0.02$}\textsuperscript{j} & $\leq 26.20$\textsuperscript{l} & 3\textsuperscript{k} \\ \hline
\end{tabular}
\vspace{10pt}

\tablebib{(a)~\cite{Bellotti2024b}\swp{;} (b)~\cite{Cristofari2023}\swp{;}  (c)~\cite{Schmitt2004}\swp{;} (d)~\cite{Donati2025}\swp{;} (e)~\cite{Stock2020}\swpt{;}  (f)~\cite{Schweitzer2019}\swp{;} (g)~\cite{Pineda2026}\swp{;} (h)~\cite{Biswas2025}\swpt{;} (i)~\cite{Donati2023}\swp{;} (j)~\cite{Cristofari2022}\swp{;} (k)~\cite{Lehmann2024}\swp{;} (l)~\cite{Wright2018}.}
\label{tab:sample_properties}
\end{table*}

\begin{table*}[]
\caption{Summary of orbital parameters for confirmed exoplanets around the stars included in the sample.}
\centering
\begin{tabular}{lccccl}\hline\hline
  \begin{tabular}[c]{@{}l@{}}\swp{Exoplanet}\end{tabular} &
  \begin{tabular}[c]{@{}l@{}}\swp{$P_{\rm orb}$} {[}d{]}\end{tabular} &
  \begin{tabular}[c]{@{}l@{}}\swp{$a$} {[}au{]}\end{tabular} &
  \begin{tabular}[c]{@{}l@{}}\swp{$a$} {[}\swpt{$R_\star$}{]}\end{tabular} &
  \begin{tabular}[c]{@{}l@{}}\swp{$e$}\end{tabular} & \swpt{Reference}\\ \hline

\swp{DS Leo b} & \swp{$6.020 \pm 0.004$}  & \swp{$0.0531 \pm 0.0006$}  & \swp{$21.5 \pm 0.4$}   & \swpt{0} \swp{(fixed)} & \cite{Carmona2025}   \\

YZ Cet b & \swp{$1.96876 \pm 0.00021$} & \swp{$0.01557 \pm 0.00052$} & \swpt{$21.3 \pm 0.7$}   & \swp{$0.00 \pm 0.12$} & \cite{Astudillo-Defru2017}  \\
YZ Cet c  & \swp{$3.06008 \pm 0.00022$}  & \swp{$0.02090 \pm  0.00070$}  & \swpt{$28.6 \pm 0.9$}   & \swp{$0.04 \pm 0.11$} & \cite{Astudillo-Defru2017}   \\
YZ Cet d & \swp{$4.65627 \pm 0.00042$}  & \swp{$0.02764 \pm 0.00093$} & \swpt{$37.8 \pm 1.2$}   & \swp{$ 0.129 \pm 0.096$} & \cite{Astudillo-Defru2017} \\

GJ 1151c & \swp{$389.70^{+5.40}_{-6.50}$} & \swp{$0.571^{+0.005}_{-0.006}$}    & \swp{$646.1^{+5.7}_{-6.8}$}  & \swpt{0} \swp{(fixed)} & \cite{Blanco-Pozo2023} \\\hline
\end{tabular}
\vspace{10pt}

\label{tab:exoplanet_orbits}
\end{table*}

\subsection{Large-scale magnetic field}
We provide a brief summary of the published ZDI maps used in this work. \swp{The average radial magnetic field strengths for all ZDI maps can be found in Table \ref{tab:sw_characteristics}.}
\begin{itemize}
\item The large-scale magnetic topology of CN Leo is an axisymmetric dipole with average unsigned radial magnetic field strength $\langle |B_r| \rangle$ varying between about $303$ and $463~{\rm G}$ during the epochs used in this work (in decimal years: 2019.4, 2020.1, 2021.2, 2022.2). The magnetic field obliquity (defined as the angle between the rotational axis and the magnetic axis of the dipole component) stayed between 43 and 52 degree during all epochs \citep[][]{Bellotti2024b}.
\item EV Lac has a non-axisymmetric poloidal field dominated by the dipolar component, which was observed to have a strong jump in $\langle |B_r| \rangle$ between 2019 and 2020 from about 58 to about \swpt{$255~{\rm G}$} (epochs: 2019.9, 2020.8, 2021.6, 2023.7). The obliquity of the field has increased during the first three epochs \citep[][]{Bellotti2024b}. \cite{Donati2025} obtained magnetic field maps for the epoch 2023.7 with three approaches (1) using circularly polarized light only (2) using unpolarized and circularly polarized light, and (3) using unpolarized, circularly polarized, and linearly polarized light. For consistency with the other maps we used the map reconstructed with circular polarization. In Appendix \ref{appendix:StokesIVQU} we investigated the differences between simulation outputs with these three maps for the same epoch.
\item \swp{DS Leo} (also known as \swp{GJ 410}) has a strong toroidal component (43 and 32 percent of total magnetic field energy in 2020.9 and 2022.1 respectively), which is axisymmetric, while the poloidal component is non-axisymmetric. $\langle |B_r| \rangle$ went from $28$ to $13~{\rm G}$ between epochs \citep[][]{Bellotti2024b}. \swp{DS Leo} has been shown to have strong differential rotation \citep[][]{Hebrard2016}
\item YZ Cet has a poloidal field, where the strength and topology changed significantly between epochs. $\langle |B_r|\rangle$ changed from $157~{\rm G}$ in 2023.9 to $220~{\rm G}$ in 2024.8. The topology changed from axisymmetric with large contribution of higher spherical harmonics modes to a dipolar, non-axisymmetric magnetic field \citep[][]{Biswas2025}.
\item \swp{GJ} 905 \swp{was observed in} 2019.6, 2020.7, and 2021.8. It has a dipole dominated magnetic field that has consistently decreased between epochs. $\langle|B_r|\rangle$ went from 98 G in 2019.6 to $48~{\rm G}$ in \swpt{2021.8}. The axisymmetry of the large-scale magnetic field has decreased significantly, which might be indicative of a polarity reversal \citep[][]{Lehmann2024}.
\item GJ 1151 has a poloidal field with significant contribution of higher spherical harmonics modes than the dipole. $\langle |B_r| \rangle$ increased from $17~{\rm G}$ to $44~{\rm G}$ during the epochs \swp{in which it was observed} (2020.4, 2021.2, 2022.2), and it became more non-axisymmetric. \swp{This is the only star in our sample where the available magnetic field maps show a polarity reversal.} \citep[][]{Lehmann2024}
\end{itemize}

\section{\swpt{Analysis and results}}\label{section:stellar_wind}

\begin{figure*}
    \sidecaption
    \includegraphics[width=12cm]{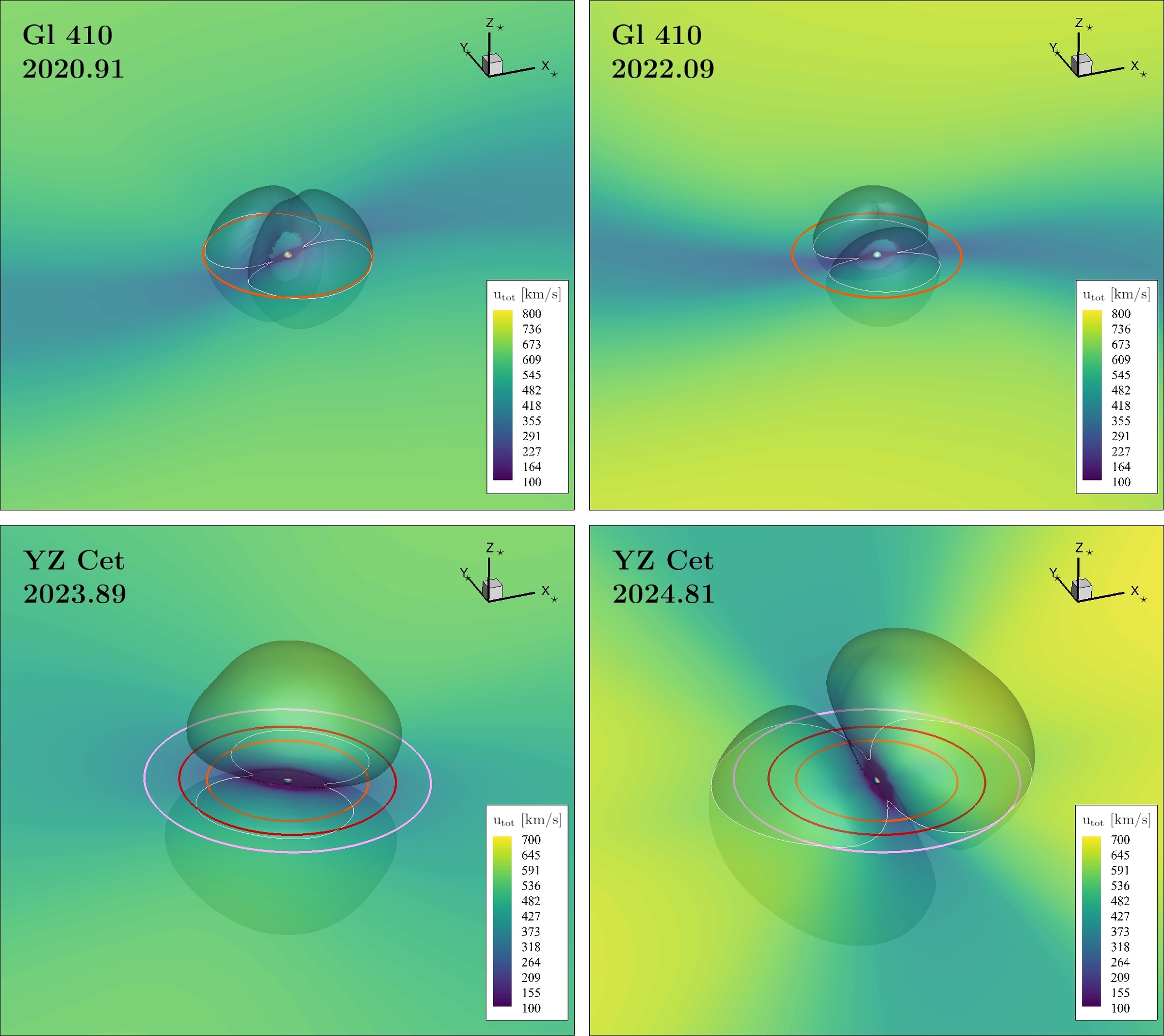}
    \caption{Three-dimensional outputs of the stellar wind simulations. The color bar represents the total stellar wind velocity. The shaded region centered around the star is the Alfvén surface and the white curve indicates its projection on the $x_*-y_*$ plane. Exoplanet orbits, which are assumed to be constrained in the equatorial plane, are also shown as concentric curves. For YZ Cet, the orange, red and pink orbit refer to planets b, c, and d (see Table \ref{tab:exoplanet_orbits}).}
    \label{fig:simulation_3d}
\end{figure*}

\subsection{\swpt{Stellar wind simulation setup}}

We simulated the stellar wind for each magnetic field map using the Space Weather Modeling Framework \citep[SWMF;][]{Toth2005, Toth2012} with the Alfvén wave solar model \citep[AWSoM;][]{Sokolov2013, vanderHolst2014} extension. The simulation numerically solves the three dimensional ideal two-temperature magnetohydrodynamic equations. In each iteration, the models are updated until convergence to a steady-state solution, where magnetic and hydrodynamic forces are in balance across the entire domain of the simulation. Even though this model was developed for the Solar wind, it has been applied to winds of cool stars \citep[e.g.][]{Evensberget2021}.

Our \swpt{\texttt{AWSoM}} model extends from the chromosphere (inner boundary) up to 100 stellar radii. The chromosphere temperature is set to $5\times10^4~K$ and number density to $2 \times 10^{5}~\text{cm}^{-3}$. In the model, the plasma is heated by dissipation of Alfvén waves. We assume a base Poynting flux $S_A$, which is proportional to \swp{the} local magnetic field strength $B$ at the inner boundary of the model. We specify it by defining the Poynting-flux-to-magnetic-field-ratio as $S_A/B = 1.1 \times 10^5~\text{erg cm}^{-2}\text{s}^{-1}\text{ G}^{-1}$, which is consistent with \swp{s}olar wind models. Finally, we set turbulence correlation length to $1.5 \times 10^{13}~\text{cm G}^{1/2}$. The parameters used in this work have been shown to reproduce \swp{s}olar wind conditions \citep[][]{Meng2015, vanderHolst2019, Sachdeva2019}. A discussion on the impact of these parameters on simulation results can be found in \cite{BoroSaikia2020}, \cite{Fionnagain2021}, and \cite{Jivani2023}.

Additional inputs to the simulation are specific to the star - stellar mass, radius, rotation period, and large-scale magnetic field. Specifically, the radial component of the magnetic field map obtained with ZDI is provided and fixed at the inner boundary of the simulation, while the transverse components are allowed to float.

\subsection{Characteristics of the \swpt{stellar} wind}
We characterize the stellar environment in each of the \swp{18} simulations by computing mass-loss rate, angular momentum-loss rate and determining the Alfvén surface. Mass-loss rate $\dot{M}$ is given by integrating mass flux over an arbitrary surface enclosing the star:
\begin{equation}\label{eq:mdot_formula}
    \dot{M} = \oint_S \rho \mathbf{u} \cdot d\mathbf{S}
\end{equation}
Where $\rho$ is the stellar wind density, $\mathbf{u}$ is the velocity vector of the stellar wind, and $d\mathbf{S}$ is a vector representation of an infinitesimal area element. In our computations, we chose a spherical surface. Once the simulation achieves steady state, we expect the mass-loss rate to be independent of the size of the sphere. Due to numerical effects, we observed small variations with the radius of the sphere used in the calculation. Thus, we sampled the mass-loss rate for 20 linearly spaced surfaces between 10 and 38 stellar radii. We found typical deviations of $\dot{M}$ between these shells to be less than 1\% of the average value. We took a similar approach for computing the angular momentum-loss rate $\dot{J}$, which is given by \citep{Vidotto2014b}:
\begin{equation}\label{eq.jdot}
    \dot{J} = \oint_{S_{\text{sph}}} \left(-\cfrac{\varpi B_\varphi B_r}{4 \pi} + \varpi u_\varphi \rho u_r\right) dS_{\text{sph}}
\end{equation}
where $\varpi = \sqrt{x^2 + y^2}$ is the cylindrical radius, subscripts $\varphi$ and $r$ indicate the azimuthal and the radial components of the velocity $\mathbf{u}$ and the magnetic field $\mathbf{B}$. \aav{Equation (\ref{eq.jdot})} is derived assuming a spherical surface centered at the star. We write $dS_{\text{sph}}$ to emphasize that, unlike in equation \ref{eq:mdot_formula}, a spherical surface is necessary. Typical deviation\swp{s} from average value of $\dot{J}$ due to choice of the spherical shell are less than 10\% of the average value. 

We also characterize the size of the Alfvén surface by computing the average Alfvén radius $R_A$. The Alfvén surface (\swp{semi-transparent region} in Fig. \ref{fig:simulation_3d}) is the region where the regime transitions from sub-Alfvénic (local wind speed $u$ is smaller than Alfvén speed $u_A$) to super-Alfvénic ($u > u_A$). The average Alfvén radius can also be used to compute the angular momentum-loss rate using analytical models \citep[][]{WeberDavis1967}. To characterize the Alfvén surface, we computed the Alfvénic Mach number given by $M_A =u /u_A$, and then identified the iso-surface where $M_A = 1.0$. Then we computed the average distance between the Alfvén surface and the center of the star (1) using an unweighted average and (2) using local mass flux as weights. The first of these approaches, which we denote as $\langle R_A\rangle$, is more relevant for star-planet interactions. That is because for the nature \swp{of} star-planet interaction\swp{s} the most relevant factor is whether the information can travel upstream (which is not possible in the super-Alfvénic regime). The mass-flux averaged Alfvén radius, which we denote as $\langle \tilde{R_A} \rangle$, is useful for analytical calculations of angular momentum-loss rate \citep[e.g.][]{WeberDavis1967} because in this case we also need to consider which regions are removing the most material, thus contributing to bigger angular momentum loss.

\swpt{In Table \ref{tab:sw_characteristics} we provide the mass-loss rate, the angular momentum-loss rate, and the average Alfvén radius for each star in each of the epochs available. We provide these values computed with two values of Poynting-flux-to-magnetic-field ratio, which is discussed further in Sect. \ref{section:poynting}.}

\begin{table*}[th!]
\caption{Stellar wind characteristics derived from simulation outputs.}
\centering
\resizebox{\textwidth}{!}{\begin{tabular}{lccccccccccc}
\hline\hline \\
Star & Epoch & \swp{$\langle|B_r|\rangle\,[\rm G]$} &
  
  \multicolumn{2}{c}{\begin{tabular}{c}\swp{$\dot{M} ~[10^{-13}~\rm M_\odot\,yr^{-1}]$} \vspace{3pt} \\  \end{tabular}} &
  \multicolumn{2}{c}{\begin{tabular}{c}$\dot{J} ~[10^{29} \rm erg]$ \vspace{3pt} \\ \end{tabular}} &
  \multicolumn{2}{c}{\begin{tabular}{c}\swpt{$\langle R_A \rangle ~[R_\star]$} \vspace{3pt}\\ \end{tabular}} &
  \multicolumn{2}{c}{\begin{tabular}{c}\swpt{$\langle \tilde{R_A}\rangle ~[ R_\star]$} \vspace{3pt}\\\end{tabular}}\vspace{2pt} \\
                          & & & $(S_A/B)_\odot$ &$(S_A/B)_{\rm reduced}$ & $(S_A/B)_\odot$ & $(S_A/B)_{\rm reduced}$ & $(S_A/B)_\odot$ & $(S_A/B)_{\rm reduced}$ & $(S_A/B)_\odot$ & $(S_A/B)_{\rm reduced}$ \vspace{3pt} \\ \hline
\multirow{4}{*}{CN Leo}                 & 2019.4 & \swp{$326$} & 0.4 & $3.1\times10^{-2}$  & 8.5   & $0.8$  & 37 & 50  & 29 & 42 \\
                         & 2020.1 & \swp{$303$} & 0.4 & $3.0\times10^{-2}$  & 8.1   & 0.8  & 36 & 51  & \swpt{29} & 43  \\
                         & 2021.2 & \swp{$463$} & 0.5 & $3.4\times10^{-2}$  & 9.4   & 0.9  & 41 & 47  & 33 & 43  \\
                         & 2022.2 & \swp{$402$} & 0.5 & $3.2\times10^{-2}$  & 9.0   & 0.9  & 39 & 47  & 32 & 43  \\
                         \noalign{\smallskip}
\multirow{4}{*}{EV Lac}  & 2019.9 & \swp{$58$} & 0.6 & $4.1\times10^{-2}$  & 66.4   & 11.9  & 24 & 44  & 16 & 31  \\
                         & 2020.8 & \swp{$255$} & 2.2 & $0.1$  & 493.0 & 112.0  & 35 & 50  & 25 & 39  \\
                         & 2021.6 & \swp{$227$} & 2.3 & 0.1  & 484.0 & 122.0  & 34 & 50  & 25 & 39  \\
                         & 2023.7 & \swp{$281$} & 1.8 & 0.1 & 351.0 & 93.6 & 30 & 52 & 22 & 38 \\
                         \noalign{\smallskip}
\multirow{2}{*}{\swp{DS Leo}}  & 2020.9 & \swp{$28$} & 1.0 & $6.9\times10^{-2}$  & 79.7  & 18.1  & 18 & 35  & 13 & 26  \\
                         & 2022.1 & \swp{$13$} & 0.5 & $2.1\times10^{-2}$  & 26.5  & 4.1  & 16 & 30  & 11 & 23  \\
                         \noalign{\smallskip}
\multirow{2}{*}{YZ Cet}  & 2023.9 & \swp{$157$} & 0.5 & $3.6\times10^{-2}$  & 1.0   & $0.1$  & 30 & 50  & 22 & 37  \\
                         & 2024.8 & \swp{$220$} & 0.6 & $4.7\times10^{-2}$  & 1.6   & $0.3$  & 34 & 49 & 22 & 37  \\
                         \noalign{\smallskip}
\multirow{3}{*}{\swp{GJ} 905}  & 2019.6 & \swp{$98$} & 0.3 & $2.3\times10^{-2}$  & 0.3   & $7.4\times10^{-2}$  & 28 & 47  & 19 & 34  \\
                         & 2020.7 & \swp{$68$} & 0.3 & $1.6\times10^{-2}$  & 0.2   & $2.6\times10^{-2}$  & 24 & 44  & 17 & 32  \\
                         & 2021.8 & \swp{$48$} & 0.2 & $1.3\times10^{-2}$  & 0.3   & $7.6\times10^{-2}$  & 22 & 41  & 16 & 31  \\
                         \noalign{\smallskip}
\multirow{3}{*}{GJ 1151} & 2020.4 & \swp{$17$} & 0.1 & $4.2\times10^{-3}$  & \swp{$4.6\times10^{-2}$}   & $4.0\times10^{-3}$  & 16 & 31  & 13 & 27  \\
                         & 2021.2 & \swp{$24$} & 0.1 & $7.0\times10^{-3}$  & \swp{$8.3\times10^{-2}$}  & $1.0\times10^{-2}$  & 18 & 35  & 13 & 26  \\
                         & 2022.2 & \swp{$44$} & 0.3 & $1.8\times10^{-2}$  & 0.2   & $4.7\times10^{-2}$  & 22 & 40  & 16 & 30 \\ \hline
\end{tabular}}
\vspace{5pt}

\tablefoot{Each row represents simulations corresponding to a single ZDI magnetic field map: one with Solar Poynting-flux-to-magnetic-field ratio $(S_A/B)_\odot = 1.1\times10^5~{\rm erg~cm^{-2}~s^{-1}~G^{-1}}$ and one with $(S_A/B)_{\rm reduced} = 1.1\times10^4~{\rm erg~cm^{-2}~s^{-1}~G^{-1}}$. Mass-loss rate ($\dot{M}$), angular momentum-loss rate ($\dot{J}$), average Alfvén radius ($\langle R_A \rangle$), and average Alfvén radius weighed by local mass flux ($\langle \tilde{R_A} \rangle$) are given.}
\label{tab:sw_characteristics}
\end{table*}

\subsection{\swpt{Temporal variability of the stellar wind}}
\begin{figure*}[h!]
\sidecaption
    \includegraphics[width=12cm]{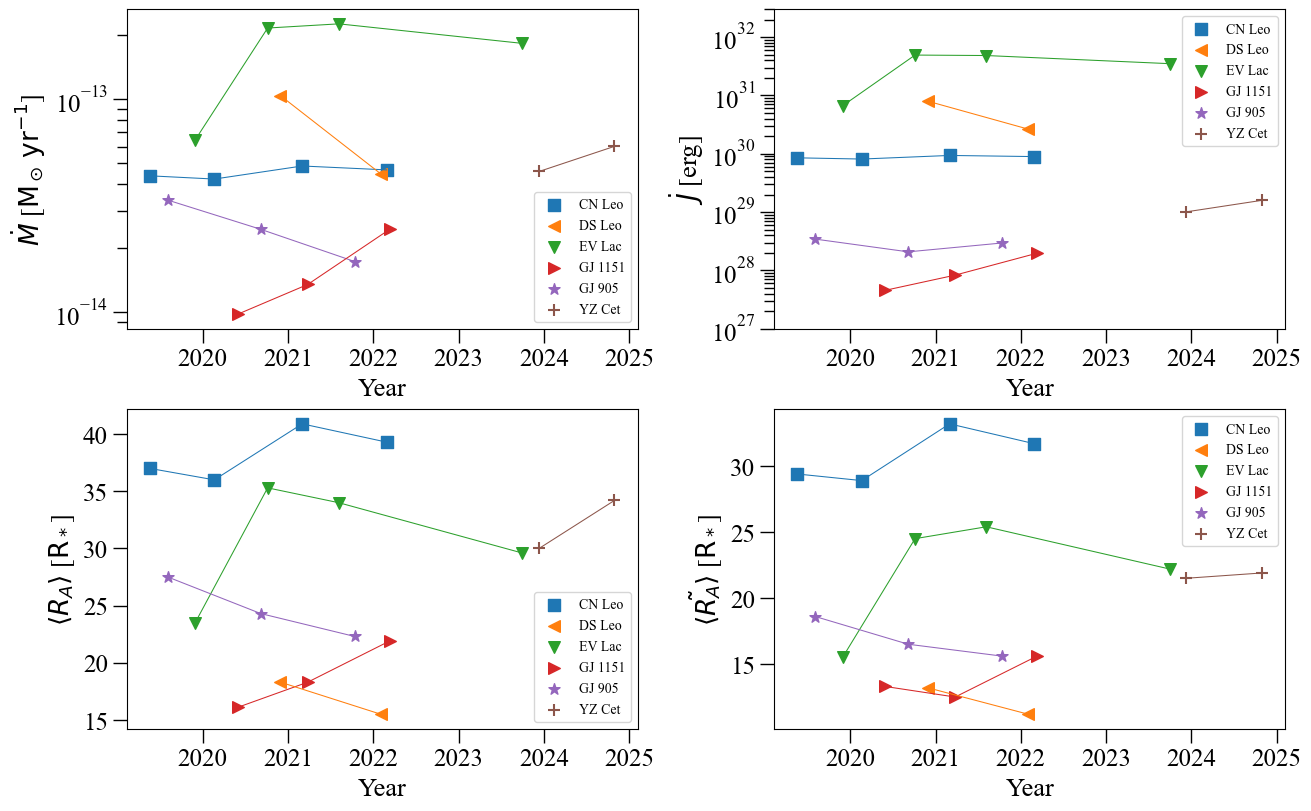}
        \caption{Variability of stellar environment properties, obtained from simulations, for stars in our sample. The top-left plot shows variability of mass-loss rate, the top-right plot shows variability of angular momentum-loss rate. The two bottom plots show variability of the average Alfvén radius, computed with (bottom-right) and without (bottom-left) weighting grid points by local mass flux.}
        \label{fig:sw_characteristics_variability}
\end{figure*}

In Fig. \ref{fig:sw_characteristics_variability} we display the mass-loss rate, angular momentum-loss rate, and average Alfvén radius determined for all simulations. Some temporal variability is observed for all stars and properties. In particular, the size of the Alfvén surface can change significantly, which is most visible for EV Lac where the average Alfvén radius (both $\langle R_A\rangle$ and $\langle \tilde{R_A}\rangle$) increased by a factor of 1.5 going from \swpt{$24~R_\star$ to $35~R_\star$} in case of $\langle R_A\rangle$ and from \swpt{$16~R_\star$ to $25~R_\star$} in case of $\langle \tilde{R_A}\rangle$. This corresponds to an increase in $\langle |B_r|\rangle$ (average large-scale radial magnetic field strength) of EV Lac by a factor of about 4.4 (from $58~{\rm G}$ to \swpt{$255~{\rm G}$}). On the opposite end we have CN Leo where the amplitude of variability of $R_A$ is \swp{approximately} 1.1, corresponding to changes in $\langle |B_r|\rangle$ by a factor of 1.5 (from $303~{\rm G}$ to \swpt{$463~{\rm G}$}). Overall, we found that for our sample $\langle R_A\rangle \propto \langle |B_r|\rangle^{0.24\pm0.02}$ provides a good approximation, which is consistent with the scaling law found \swp{for Sun-like stars} in \cite{Evensberget2022}, and comparable to those in \cite{Evensberget2021, Evensberget2023} where the exponent was higher ($0.412\pm0.035$ and $0.33\pm0.04$ respectively). The method used for computation of the Alfvén surface also shows significant impact on the result. It is expected that $\langle \tilde{R_A} \rangle < \langle R_A \rangle$ since the first of these estimates weighs by local mass flux. Therefore regions of the Alfvén surface near the star, where wind density is higher, will be weighted more when averaging than the distant regions. We also see that the trends between the two methods differ - for example GJ 1151 has a monotonic increase in Alfvén radius, while the mass flux weighted Alfvén radius first dips, and then increases. The opposite is observed for EV Lac. In case of YZ Cet the Alfvén radius changes significantly with one method while changing less prominently with the other. The difference in trends between the two estimators reflects the complexity of the geometry of the Alfvén surface (dictated by the magnetic field geometry).

The mass-loss rate remains roughly within an order of magnitude for all stars, ranging between \swp{$1.0 \times 10^{-14}~M_\odot/{\rm yr}$} for GJ 1151, and \swp{$2.3\times10^{-13}~M_\odot/{\rm yr}$} for EV Lac. In case of CN Leo we found negligible variations in $\dot{M}$ over time. \swpt{The biggest change is observed for EV Lac, where $\dot{M}$ differs by a factor of 3.8 between the epochs with minimum and maximum values of mass-loss rate.}

We found angular momentum-loss rates between \swp{$4.6\times10^{27}~{\rm erg}$} for GJ 1151 and \swp{$4.9\times10^{31}~{\rm erg}$} for EV Lac. The large spread of values agrees with the analytical expression of the Weber-Davis model (see Appendix \ref{appendix:poynting_WD}). The changes in $\dot{J}$ for individual stars are more pronounced than the variability of the mass-loss rate, varying by \swpt{up to factor 7.4 for} EV Lac. CN Leo shows the smallest variation in angular momentum-loss rate, while \swp{GJ} 905, YZ Cet, \swp{DS Leo}, and \swpt{GJ 1151} show variations in between these two cases.

\cite{Smith2026} investigated the variability of stellar wind properties of AD Leo, an M3.5 star, using \swpt{\texttt{AWSoM}} simulations of the stellar wind based on five ZDI maps over 4 years. They found variability of $\dot{M}$ by up to factor 4.3, between $1.6 \times 10^{-13}~M_\odot/{\rm yr}$ and $6.8 \times 10^{-13}~M_\odot/{\rm yr}$, which is slightly higher than the variability we determined for stars in our sample. A similar study was conducted for F-type star $\chi$ Dra  by \cite{Marsden2023}, also using the \swpt{\texttt{AWSoM}} model. They used four ZDI maps obtained over 5 years to investigate the variability of stellar winds . They found $\dot{M}$ varying by up to 1.3 times, between $7.4 \times 10^{-14}~M_\odot/{\rm yr}$ and $9.9 \times 10^{-14}~M_\odot/{\rm yr}$, and $\dot{J}$ by up to factor 1.6, between $5.4 \times 10^{30}~\rm erg$ and $8.6 \times 10^{30}~\rm erg$. The variability found for $\chi$ Dra is consistent with what we determined for M dwarfs in our sample.

\subsection{Environment at planetary orbits}\label{section:exoplanet_sw}

\begin{figure*}[h!]
    \centering
    \includegraphics[width=1\linewidth]{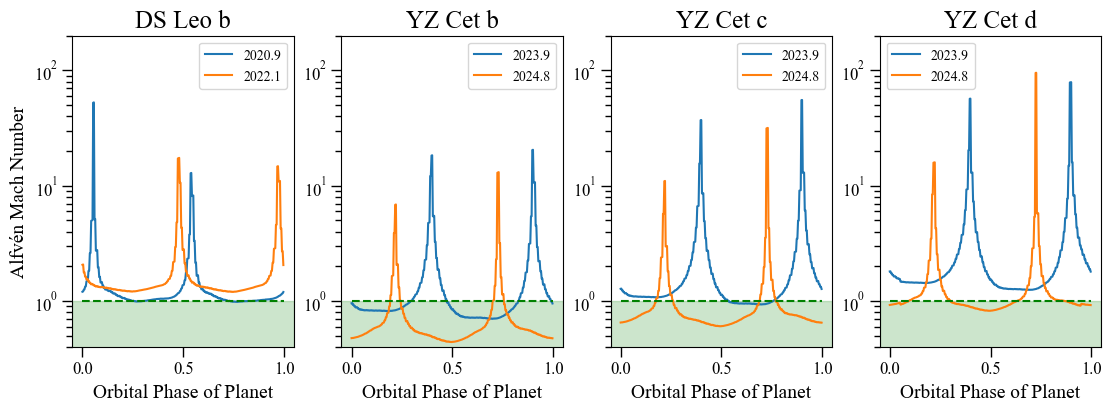}
    \caption{Mach number variation through the orbit of confirmed exoplanets around \swp{DS Leo} and YZ Cet. The colors of the curves correspond to different observational epochs. The green shaded area indicates a sub-Alfvénic regime.}
    \label{fig:mach_number_evolution}
\end{figure*}

For the stars that have confirmed exoplanets, namely \swp{DS Leo}, GJ 1151, and YZ Cet (see Table \ref{tab:exoplanet_orbits}), we analyzed the properties of the stellar wind at the orbit. Since these exoplanets are only detected with radial velocity method, the inclinations of their orbits are not constrained. Thus, we assume all the companions to orbit in the equatorial plane. We sample the stellar wind parameters along such orbits, and compute the Alfvénic mach number in the planetary restframe at each point, \aav{allowing} us to determine whether an orbit is super-\swp{Alfvénic} ($M_A > 1$) or sub-\swp{Alfvénic} ($M_A < 1$). The results are displayed in Fig. \ref{fig:mach_number_evolution}, with the exception of GJ 1151 c which will remain super-Alfvénic at all times due to its large distance from the host star. We observe significant temporal variability in the fraction of the orbits being super or sub-\swp{Alfvénic}. For \swp{DS Leo b} we found 15\% of the orbit was in the sub-Alfvénic regime in the 2020.9 epoch, while the orbit was completely super-Alfvénic in 2022.1.

Similarly, the orbit of YZ Cet d was fully super-Alfvénic in 2023.9, while crossing inside the Alfvén surface in 2024.8 and spending 55\% of the orbit in the sub-Alfvénic regime. For YZ Cet b, 66\% and 89\% of the orbit was sub-Alfvénic in 2023.9 and 2024.8 respectively. For YZ Cet c, 22\% and 83\% of the orbit was sub-\swp{Alfvénic} in the respective epochs. These results show that variations of the large-scale field can cause the nature of the star-planet interactions to change significantly on timescales of a few months, affecting the possibility of detection of SPI signatures such as e.g. \swp{modulation in chromospheric activity indicators \citep[e.g., ][]{Shkolnik2003} or} planet-induced flaring \citep[][]{Ilin2025}. In particular, magnetic field variability can explain why some potential SPI detections are not repeatable \citep[e.g.][]{Shkolnik2008, Scandariato2013}. Therefore, simultaneous multi-wavelength observations of planetary systems are essential to understanding SPIs,  in agreement with previous papers in the field that utilize a multi-wavelength approach or advocate a similar conclusion \citep[e.g.][]{Bourrier2020, Vidotto2025, Callingham2025, Bellotti2025a}.

\begin{table*}[th!]
\centering
\caption{Simulated stellar wind parameters at orbits of known exoplanets orbiting M dwarfs in the sample.}
\resizebox{\textwidth}{!}{\begin{tabular}{lcccccccc}
\hline\hline
Exoplanet &
  Epoch &
  \begin{tabular}[c]{@{}l@{}}\swp{$\rho$} $[10^{-20} ~{\rm g~cm^{-3}}]$ \end{tabular} &
  \begin{tabular}[c]{@{}l@{}}\swp{$u$} $[{\rm km~s^{-1}}]$\end{tabular} &
  \begin{tabular}[c]{@{}l@{}}\swp{$B$} $[\rm m G]$\end{tabular} &
  \begin{tabular}[c]{@{}l@{}}\swp{$T_i$} $[{\rm M K}]$\end{tabular} &
  \begin{tabular}[c]{@{}l@{}}\swp{$T_e$} $[{\rm MK}]$\end{tabular} &
  \begin{tabular}[c]{@{}l@{}}\swp{$P_{\rm ram}$} $[10^{-5}~{\rm dyne~cm^{-2}}]$\end{tabular} &
  \begin{tabular}[c]{@{}l@{}}Sub-Alfvénic \\ \swpt{orbit fraction}\end{tabular}\\ \hline

\multirow{2}{*}{\swp{\vspace{-6pt}DS Leo b}}
  & 2020.9
  & \begin{tabular}[c]{@{}c@{}}1.8\\[-2pt]\swp{$[0.8,\,9.8]$}\end{tabular}
  & \begin{tabular}[c]{@{}c@{}}449\\[-2pt]\swp{$[324,\,515]$}\end{tabular}
  & \begin{tabular}[c]{@{}c@{}}16\\[-2pt]\swp{$[0.7,\,18]$}\end{tabular}
  & \begin{tabular}[c]{@{}c@{}}0.34\\[-2pt]\swp{$[0.18,\,0.48]$}\end{tabular}
  & \begin{tabular}[c]{@{}c@{}}1.59\\[-2pt]\swp{$[1.13,\,1.92]$}\end{tabular}
  & \begin{tabular}[c]{@{}c@{}}3.2\\[-2pt]\swp{$[2.2,\,12]$}\end{tabular}
  & 15\% \\

  & 2022.1
  & \begin{tabular}[c]{@{}c@{}}$0.8$\\[-2pt]\swpt{$[0.3,\,3.8]$}\end{tabular}
  & \begin{tabular}[c]{@{}c@{}}490\\[-2pt]\swp{$[306,\,576]$}\end{tabular}
  & \begin{tabular}[c]{@{}c@{}}9\\[-2pt]\swp{$[1.2,\,10]$}\end{tabular}
  & \begin{tabular}[c]{@{}c@{}}0.59\\[-2pt]\swp{$[0.22,\,0.84]$}\end{tabular}
  & \begin{tabular}[c]{@{}c@{}}1.62\\[-2pt]\swp{$[1.20,\,1.90]$}\end{tabular}
  & \begin{tabular}[c]{@{}c@{}}1.5\\[-2pt]\swp{$[1.1,\,4.0]$}\end{tabular}
  & 0\% \\
  \noalign{\smallskip}

\multirow{2}{*}{\vspace{-6pt}YZ Cet b}
  & 2023.9
  & \begin{tabular}[c]{@{}c@{}}18.6\\[-2pt]\swp{$[7.3,\,52.5]$}\end{tabular}
  & \begin{tabular}[c]{@{}c@{}}379\\[-2pt]\swp{$[284,\,524]$}\end{tabular}
  & \begin{tabular}[c]{@{}c@{}}51\\[-2pt]\swp{$[3.8,\,61]$}\end{tabular}
  & \begin{tabular}[c]{@{}c@{}}0.45\\[-2pt]\swp{$[0.23,\,1.1]$}\end{tabular}
  & \begin{tabular}[c]{@{}c@{}}1.67\\[-2pt]\swp{$[1.48,\,1.93]$}\end{tabular}
  & \begin{tabular}[c]{@{}c@{}}23.8\\[-2pt]\swp{$[14.5,\,48.7]$}\end{tabular}
  & 66\% \\

  & 2024.8
  & \begin{tabular}[c]{@{}c@{}}13.8\\[-2pt]\swp{$[5.5,\,81.5]$}\end{tabular}
  & \begin{tabular}[c]{@{}c@{}}466\\[-2pt]\swp{$[254,\,552]$}\end{tabular}
  & \begin{tabular}[c]{@{}c@{}}84\\[-2pt]\swp{$[6.0,\,95]$}\end{tabular}
  & \begin{tabular}[c]{@{}c@{}}0.36\\[-2pt]\swp{$[0.20,\,1.33]$}\end{tabular}
  & \begin{tabular}[c]{@{}c@{}}1.94\\[-2pt]\swpt{$[1.35,\,2.13]$}\end{tabular}
  & \begin{tabular}[c]{@{}c@{}}22.6\\[-2pt]\swp{$[13.7,\,65.2]$}\end{tabular}
  & 89\% \\
  \noalign{\smallskip}

\multirow{2}{*}{\vspace{-6pt}YZ Cet c}
  & 2023.9
  & \begin{tabular}[c]{@{}c@{}}9.7\\[-2pt]\swp{$[4.0,\,26.7]$}\end{tabular}
  & \begin{tabular}[c]{@{}c@{}}412\\[-2pt]\swp{$[336,\,482]$}\end{tabular}
  & \begin{tabular}[c]{@{}c@{}}30\\[-2pt]\swp{$[1.2,\,36]$}\end{tabular}
  & \begin{tabular}[c]{@{}c@{}}0.35\\[-2pt]\swp{$[0.20,\,0.76]$}\end{tabular}
  & \begin{tabular}[c]{@{}c@{}}1.51\\[-2pt]\swp{$[1.38,\,1.66]$}\end{tabular}
  & \begin{tabular}[c]{@{}c@{}}15.0\\[-2pt]\swp{$[8.8,\,33.0]$}\end{tabular}
  & 22\% \\

  & 2024.8
  & \begin{tabular}[c]{@{}c@{}}7.2\\[-2pt]\swp{$[2.8,\,40.7]$}\end{tabular}
  & \begin{tabular}[c]{@{}c@{}}498\\[-2pt]\swp{$[309,\,581]$}\end{tabular}
  & \begin{tabular}[c]{@{}c@{}}48\\[-2pt]\swp{$[2.3,\,53]$}\end{tabular}
  & \begin{tabular}[c]{@{}c@{}}0.32\\[-2pt]\swp{$[0.19,\,0.91]$}\end{tabular}
  & \begin{tabular}[c]{@{}c@{}}1.81\\[-2pt]\swp{$[1.56,\,2.02]$}\end{tabular}
  & \begin{tabular}[c]{@{}c@{}}13.7\\[-2pt]\swp{$[8.1,\,44.0]$}\end{tabular}
  & 83\% \\
  \noalign{\smallskip}

\multirow{2}{*}{\vspace{-6pt}YZ Cet d}
  & 2023.9
  & \begin{tabular}[c]{@{}c@{}}5.2\\[-2pt]\swp{$[2.2,\,13.8]$}\end{tabular}
  & \begin{tabular}[c]{@{}c@{}}439\\[-2pt]\swp{$[376,\,505]$}\end{tabular}
  & \begin{tabular}[c]{@{}c@{}}17\\[-2pt]\swp{$[0.6,\,21]$}\end{tabular}
  & \begin{tabular}[c]{@{}c@{}}0.27\\[-2pt]\swp{$[0.18,\,0.52]$}\end{tabular}
  & \begin{tabular}[c]{@{}c@{}}1.34\\[-2pt]\swp{$[1.25,\,1.45]$}\end{tabular}
  & \begin{tabular}[c]{@{}c@{}}9.3\\[-2pt]\swp{$[5.4,\,20.7]$}\end{tabular}
  & 0\% \\

  & 2024.8
  & \begin{tabular}[c]{@{}c@{}}3.8\\[-2pt]\swp{$[1.5,\,20.5]$}\end{tabular}
  & \begin{tabular}[c]{@{}c@{}}527\\[-2pt]\swp{$[351,\,607]$}\end{tabular}
  & \begin{tabular}[c]{@{}c@{}}27\\[-2pt]\swp{$[0.6,\,30.0]$}\end{tabular}
  & \begin{tabular}[c]{@{}c@{}}0.28\\[-2pt]\swp{$[0.18,\,0.61]$}\end{tabular}
  & \begin{tabular}[c]{@{}c@{}}1.66\\[-2pt]\swp{$[1.35,\,1.88]$}\end{tabular}
  & \begin{tabular}[c]{@{}c@{}}8.2\\[-2pt]\swp{$[4.8,\,27.2]$}\end{tabular}
  & 55\% \\
  \noalign{\smallskip}

\multirow{3}{*}{\vspace{-18pt}GJ 1151 c}
  & 2020.4
  & \begin{tabular}[c]{@{}c@{}}$3.9 \times 10^{-3}$\\[-2pt]\swp{$[3.3,\,4.4]\times10^{-3}$}\end{tabular}
  & \begin{tabular}[c]{@{}c@{}}392\\[-2pt]\swp{$[384,\,404]$}\end{tabular}
  & \begin{tabular}[c]{@{}c@{}}$4 \times 10^{-3}$\\[-2pt]\swp{$[0.3,\,6.3]\times10^{-3}$}\end{tabular}
  & \begin{tabular}[c]{@{}c@{}}\swp{$4.1\times10^{-3}$}\\[-2pt]\swp{$[4.1,\,4.1]\times10^{-3}$}\end{tabular}
  & \begin{tabular}[c]{@{}c@{}}\swp{$3.6 \times 10^{-2}$}\\[-2pt]\swp{$[3.3,\,3.9]\times 10^{-2}$}\end{tabular}
  & \begin{tabular}[c]{@{}c@{}}$6.2 \times 10^{-3}$\\[-2pt]\swp{$[5.5,\,6.7]\times10^{-3}$}\end{tabular}
  & 0\% \\

  & 2021.2
  & \begin{tabular}[c]{@{}c@{}}$2.4 \times 10^{-3}$\\[-2pt]\swp{$[1.0,\,6.0]\times10^{-3}$}\end{tabular}
  & \begin{tabular}[c]{@{}c@{}}502\\[-2pt]\swp{$[385,\,596]$}\end{tabular}
  & \begin{tabular}[c]{@{}c@{}}$16 \times 10^{-3}$\\[-2pt]\swp{$[4.1,\,19.7]\times10^{-3}$}\end{tabular}
  & \begin{tabular}[c]{@{}c@{}}\swp{$9.4\times10^{-3}$}\\[-2pt]\swp{$[4.1,\,16]\times10^{-3}$}\end{tabular}
  & \begin{tabular}[c]{@{}c@{}}\swp{$5.2 \times 10^{-2}$}\\[-2pt]\swp{$[3.5,\,6.3]\times 10^{-2}$}\end{tabular}
  & \begin{tabular}[c]{@{}c@{}}$5.1 \times 10^{-3}$\\[-2pt]\swp{$[3.6,\,9.3]\times10^{-3}$}\end{tabular}
  & 0\% \\

  & 2022.2
  & \begin{tabular}[c]{@{}c@{}}$3.2 \times 10^{-3}$\\[-2pt]\swp{$[1.3,\,12]\times10^{-3}$}\end{tabular}
  & \begin{tabular}[c]{@{}c@{}}557\\[-2pt]\swp{$[393,\,647]$}\end{tabular}
  & \begin{tabular}[c]{@{}c@{}}$28 \times 10^{-3}$\\[-2pt]\swp{$[12,\,32]\times10^{-3}$}\end{tabular}
  & \begin{tabular}[c]{@{}c@{}}\swp{$12 \times 10^{-3}$}\\[-2pt]\swp{$[4.1,\,19]\times 10^{-3}$}\end{tabular}
  & \begin{tabular}[c]{@{}c@{}}\swp{$7.0\times10^{-2}$}\\[-2pt]\swp{$[3.5,\,9.2] \times 10^{-2}$}\end{tabular}
  & \begin{tabular}[c]{@{}c@{}}$7.7 \times 10^{-3}$\\[-2pt]\swp{$[5.1,\,19]\times10^{-3}$}\end{tabular}
  & 0\% \\ \hline

\end{tabular}}
\vspace{10pt}

\tablefoot{The reported quantities are average values through an orbit of the exoplanets, with the exception of sub-Alfvénic orbit fraction, which represents how much of the exoplanet's orbit falls within the Alfvén surface. \swp{Below each quantity that has been averaged we provide a minimum and maximum value over the orbit in square brackets.} Values for GJ 1151 c were extrapolated from the simulation domain as described in the text.}
\label{tab:sw_at_orbits}
\end{table*}

In Table \ref{tab:sw_at_orbits} we provide selected properties of the stellar wind at the orbit of corresponding exoplanets, averaged over a complete revolution around the host star. GJ 1151 c is located outside of the stellar wind simulation domain so we were unable to sample the orbit directly. Instead, we extrapolated stellar wind properties following \cite{Carolan2019}. We assumed that the wind density scales like $r^{-2}$ (where $r$ is the distance from the host star), the temperature scales according to adiabatic expansion, i.e. like $r^{-4/3}$, the radial magnetic field scales with $r^{-2}$, the azimuthal magnetic field scales with $r^{-1}$, and the meridional magnetic field is negligible at far distances. Stellar wind speed reaches terminal velocity, so we assumed that it stays constant beyond the simulation domain. Ram pressure scales with $r^{-2}$. 

For \swp{DS Leo b}, we can see variability between epochs by a factor of about 2 of all properties listed with the exception of wind speed and electron temperature, which stayed roughly the same in the two epochs. For exoplanets orbiting YZ Cet, we see variations by factor of 1.4 at most in case of density. The magnetic field strength varies by about factor 1.6. The variations in wind speed balance out the density variations such that the ram pressure remains roughly constant. Variations in temperature are also small. The exoplanet GJ 1151c experienced temporal variability in wind properties by factor of about 1.6 in density, 1.4 in wind speed, up to 7 in magnetic field strength, and 1.5 in ram pressure. The variations in magnetic field strength at the orbit, which significantly exceed the magnitude of variations of the average large-scale magnetic field strength (between \swpt{$17$ and $44~{\rm G}$}), are likely caused by variations in the obliquity of the magnetic field \citep[between 5 and 56 degrees - ][]{Lehmann2024}. 

Finally, we placed our findings in the context of the Solar System. \cite{Bagenal2013} reported Solar wind density at the orbit of Earth to be about $1.3 \times 10^{-23}~{\rm g~cm^{-3}}$ - similar to our results for GJ 1151, which we determined to be between $2.4 \times 10^{-23}~{\rm g~cm^{-3}}$ and $3.9 \times 10^{-23}~{\rm g~cm^{-3}}$. \swp{DS Leo b}, and the exoplanets around YZ Cet experience stellar wind density that is 3 to 4 orders of magnitude higher. In fact, exoplanets orbiting these stars likely experience quiescent stellar wind conditions that are comparable to coronal mass ejections in the Solar System. For example, a \swp{coronal mass ejection} that impacted Mercury in 2011 was determined to have density of approximately $10^{-22}~{\rm g~cm^{-3}}$, magnetic field strength of approximately $1~\rm mG$, and wind speed of $450~{\rm km~s^{-1}}$ near the orbit of Mercury \citep[][]{Slavin2014,Exner2018}. This velocity is comparable to what we found for exoplanets around \swp{DS Leo} and YZ Cet assuming Solar $S_A/B$, while the magnetic field and density of the \swp{coronal mass ejection} that hit Mercury is about 1 to 2 order of magnitude lower (Table \ref{tab:sw_at_orbits}).

\section{Discussion}\label{section:discussion}

\subsection{Poynting-flux-to-magnetic-field \swpt{r}atio}\label{section:poynting}
\begin{figure*}[ht!]
    \sidecaption
    \includegraphics[width=12cm]{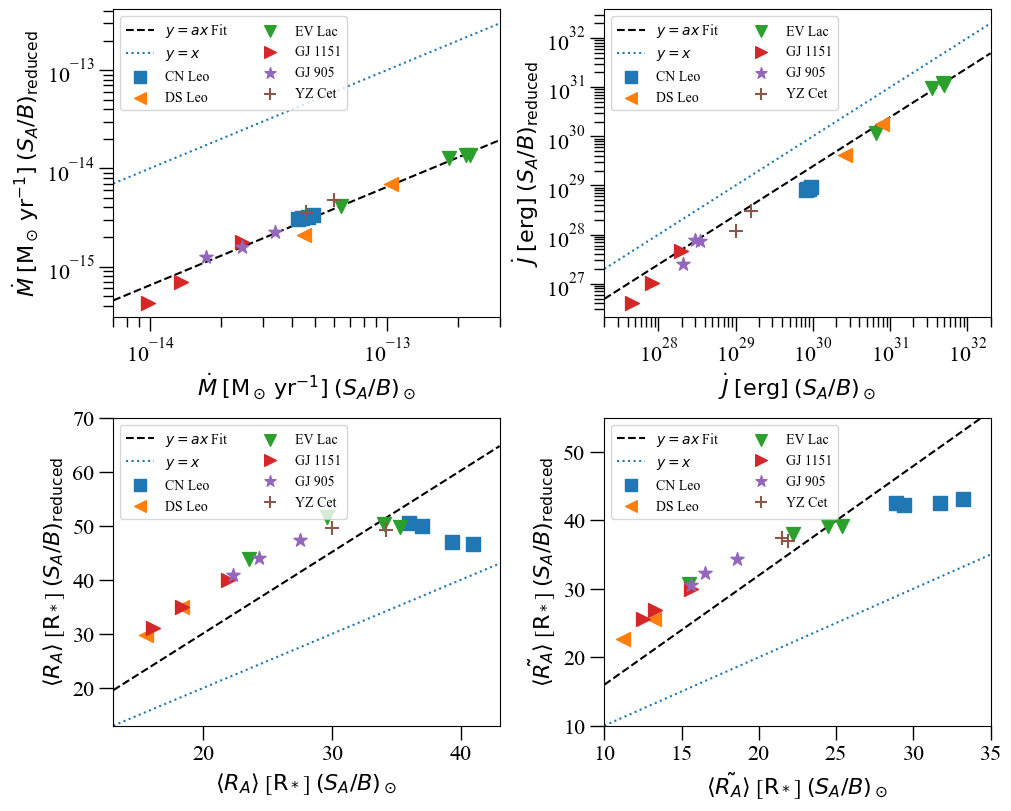}
    \caption{Comparison of stellar wind characteristics  derived with different assumptions of Poynting-flux-to-magnetic-field ratio ($S_A/B$). On the $x$ axis we provide a value obtained with the Solar value \swp{$(S_A/B)_\odot$}. On the $y$ axis we provide a value obtained with $S_A/B$ divided by 10. The top-left plot shows mass-loss rate (in Solar masses per year), the top-right plot shows angular momentum-loss rate (in erg). The two bottom plots show variability of the average Alfvén radius, computed with (bottom-right) and without (bottom-left) weighting grid points by local mass flux. In each panel, a function fit $y = ax$ is indicated with dashed lines and a one-to-one line is shown with blue dotted lines.}
    \label{fig:Poynting_Effect}
\end{figure*}

\cite{BoroSaikia2020} demonstrated that the output of the \swpt{\texttt{AWSoM}} model is strongly dependent on the Poynting-flux-to-magnetic-field ratio $S_A/B$ by investigating the Sun and Sun-like stars. Furthermore, \cite{Vidotto2023} argued that a value of $S_A/B$ that is an order of magnitude lower than the Solar value might be more suitable for reproducing the stellar wind environment around M dwarfs. To investigate the impact of $S_A/B$ on our results, we re-run all simulations with $S_A/B = 1.1 \times 10^4~\text{erg cm}^{-2}\text{s}^{-1}\text{ G}^{-1}$ (10 times lower than the Solar value). We then computed all of the stellar wind characteristics and compared them against the Solar value Poynting-flux-to-magnetic-field ratio outputs. We show the comparison of the two sets of simulations in Fig. \ref{fig:Poynting_Effect} where the $y$ axis represents the outputs corresponding to $S_A/B = 1.1 \times 10^4~\text{erg cm}^{-2}\text{s}^{-1}\text{ G}^{-1}$\swp{, denoted with $(S_A/B)_{\rm reduced}$,} and the $x$ axis corresponds to $S_A/B = 1.1 \times 10^5~\text{erg cm}^{-2}\text{s}^{-1}\text{ G}^{-1}$\swp{, denoted with $(S_A/B)_\odot$}.

Reducing the Poynting-flux-to-magnetic-field ratio $S_A/B$ decreases mass loss and angular momentum-loss rates, which is consistent with \cite{BoroSaikia2020}, and increases the average Alfvén radius. We fitted a function $y = a x$ relating the two simulation sets, where $x$ is a stellar wind parameter ($\dot{M}$, $\dot{J}$, $\langle R_A\rangle$, $\langle \tilde{R_A} \rangle$) computed with Solar $S_A/B$, $y$ is the same parameter but computed with $S_A/B$ reduced by 10, and $a$ is a coefficient to be fitted (Figure \ref{fig:Poynting_Effect}). \swp{\cite{Kavanagh2021} investigated the impact of $S_A/B$ on the mass-loss rate using an \swpt{\texttt{AWSoM}} model for Prox Cen and found that $\dot{M} \propto (S_A/B)^{1.06}$, implying that $\dot{M}[(S_A/B)_{\rm reduced}]/\dot{M}[(S_A/B)_{\odot}] =(10)^{-1.06} \simeq 0.087$. They also fitted a power law to mass-loss rates obtained by \cite{BoroSaikia2020} finding $\dot{M} \propto (S_A/B)^{1.37}$, which is equivalent to  $\dot{M}[(S_A/B)_{\rm reduced}]/\dot{M}[(S_A/B)_{\odot}] =(10)^{-1.37} \simeq 0.043$. 
We can compare our analysis to their results where} we found \swpt{$a = 0.063 = 10^{-1.20}$}, meaning that reducing Poynting flux by 10 yields a mass-loss rate on average \swpt{$0.063^{-1}\simeq 15.87$}  times lower. These results are consistent with \cite{Kavanagh2021}. For the angular momentum-loss rate we found \swp{$a = 0.245 = 10^{-0.61}$}. Following \cite{Kavanagh2021}, we performed a power law fit relating $S_A/B$ to $\dot{J}$ from the work of \cite{BoroSaikia2020}. We used their estimates with maximum spherical harmonics order $l_{\rm max}=10$ as these best match ZDI maps used in this work. Their data yields $\dot{J} \propto (S_A/B)^{0.72}$ during the Solar minimum\swp{, implying that $\dot{J}[(S_A/B)_{\rm reduced}]/\dot{J}[(S_A/B)_{\odot}] = 10^{-0.72} \simeq 0.191$,} and $\dot{J} \propto (S_A/B)^{1.08}$ during the Solar maximum\swp{, equivalent to $\dot{J}[(S_A/B)_{\rm reduced}]/\dot{J}[(S_A/B)_{\odot}] = 10^{-1.08} \simeq 0.083$}. Both scalings are stronger than our findings, placing our estimate closer to their estimates for the scaling during Solar minimum. Finally, we estimated best fits of \swp{$a = 1.479 = 10^{0.17}$} for $\langle R_A \rangle$ and \swp{$a = 1.585 = 10^{0.2}$} for $\langle \tilde{R_A} \rangle$. Figure \ref{fig:Poynting_Effect} shows that, unlike for $\dot{M}$ and $\dot{J}$, a $y=ax$ function does not represent the trend well for the Alfvén radius and a more complex relation with $S_A/B$ is needed to explain the trend.

The temporal variations appear monotonic (with the exception of $\langle R_A \rangle$ and $\langle \tilde{R_A} \rangle$) regardless of the choice of $S_A/B$, demonstrating that the choice of this free parameter does not alter our conclusions on temporal variations trends \swp{of stellar wind properties} significantly. Individual simulation outputs are affected by over an order magnitude. For completeness, in \swpt{Table \ref{tab:sw_characteristics}} we provide stellar wind characteristics computed with both $S_A/B$ values. \swp{In Table \ref{tab:sw_at_orbits_div10}} we provide a version of Table \ref{tab:sw_at_orbits} (conditions at exoplanet orbits) computed with the reduced $S_A/B$ value. \swp{It is worth noting that the picture of the variability of the Alfvénic regime at exoplanet orbit changes qualitatively as a result of reducing $S_A/B$. With $S_A/B = 1.1 \times 10^5~\text{erg cm}^{-2}\text{s}^{-1}\text{ G}^{-1}$ (Solar value) we obtained much higher variations in the percentage of the orbit spent in the sub-Alfvénic regime - up to $61\%$ point difference (see Table \ref{tab:sw_at_orbits}). Whereas with $S_A/B = 1.1 \times 10^4~\text{erg cm}^{-2}\text{s}^{-1}\text{ G}^{-1}$ exoplanets around YZ Cet and DS Leo remain predominantly sub-Alfvénic in both epochs, with the biggest difference in percentage of the orbit in the sub-Alfvénic regime being $17\%$ points for YZ Cet d (see Table \ref{tab:sw_at_orbits_div10}).}

\begin{table*}[th!]
\centering
\caption{\aav{Same as Table \ref{tab:sw_at_orbits}, but for  $S_A/B = 1.1\times10^4~{\rm erg~cm^{-2}~s^{-1}~G^{-1}}$}.}
\resizebox{\textwidth}{!}{\begin{tabular}{lcccccccc}
\hline\hline
Exoplanet &
  Epoch &
  \begin{tabular}[c]{@{}l@{}}\swp{$\rho$} $[10^{-20} ~{\rm g~cm^{-3}}]$ \end{tabular} &
  \begin{tabular}[c]{@{}l@{}}\swp{$u$} $[{\rm km~s^{-1}}]$\end{tabular} &
  \begin{tabular}[c]{@{}l@{}}\swp{$B$} $[\rm m G]$\end{tabular} &
  \begin{tabular}[c]{@{}l@{}}\swp{$T_i$} $[{\rm M K}]$\end{tabular} &
  \begin{tabular}[c]{@{}l@{}}\swp{$T_e$} $[{\rm MK}]$\end{tabular} &
  \begin{tabular}[c]{@{}l@{}}\swp{$P_{\rm ram}$} $[10^{-5}~{\rm dyne~cm^{-2}}]$\end{tabular} &
  \begin{tabular}[c]{@{}l@{}}Sub-Alfvénic \\ \swpt{orbit fraction}\end{tabular}\\ \hline

\multirow{2}{*}{\swp{\vspace{-6pt}DS Leo b}}
  & 2020.9
  & \begin{tabular}[c]{@{}c@{}}0.2\\[-2pt]\swp{$[0.1,\,0.9]$}\end{tabular}
  & \begin{tabular}[c]{@{}c@{}}308\\[-2pt]\swp{$[197,\,341]$}\end{tabular}
  & \begin{tabular}[c]{@{}c@{}}8\\[-2pt]\swp{$[0.9,\,9.3]$}\end{tabular}
  & \begin{tabular}[c]{@{}c@{}}0.26\\[-2pt]\swp{$[0.15,\,0.66]$}\end{tabular}
  & \begin{tabular}[c]{@{}c@{}}1.33\\[-2pt]\swp{$[1.23,\,1.53]$}\end{tabular}
  & \begin{tabular}[c]{@{}c@{}}0.2\\[-2pt]\swp{$[0.1,\,0.4]$}\end{tabular}
  & 92\% \\

  & 2022.1
  & \begin{tabular}[c]{@{}c@{}}$4.8 \times 10^{-2}$\\[-2pt]\swp{$[2.5,\,27]\times10^{-2}$}\end{tabular}
  & \begin{tabular}[c]{@{}c@{}}371\\[-2pt]\swp{$[200,\,417]$}\end{tabular}
  & \begin{tabular}[c]{@{}c@{}}4\\[-2pt]\swp{$[0.4,\,4.5]$}\end{tabular}
  & \begin{tabular}[c]{@{}c@{}}0.60\\[-2pt]\swp{$[0.32,\,0.82]$}\end{tabular}
  & \begin{tabular}[c]{@{}c@{}}1.37\\[-2pt]\swp{$[1.32,\,1.52]$}\end{tabular}
  & \begin{tabular}[c]{@{}c@{}}$5.5 \times 10^{-2}$\\[-2pt]\swp{$[4.5,\,12]\times10^{-2}$}\end{tabular}
  & 92\% \\
  \noalign{\smallskip}

\multirow{2}{*}{\vspace{-6pt}YZ Cet b}
  & 2023.9
  & \begin{tabular}[c]{@{}c@{}}2.8\\[-2pt]\swp{$[1.2,\,7.7]$}\end{tabular}
  & \begin{tabular}[c]{@{}c@{}}211\\[-2pt]\swp{$[77,\,294]$}\end{tabular}
  & \begin{tabular}[c]{@{}c@{}}24\\[-2pt]\swp{$[4,\,29]$}\end{tabular}
  & \begin{tabular}[c]{@{}c@{}}0.59\\[-2pt]\swp{$[0.17,\,1.5]$}\end{tabular}
  & \begin{tabular}[c]{@{}c@{}}1.50\\[-2pt]\swp{$[1.31,\,1.72]$}\end{tabular}
  & \begin{tabular}[c]{@{}c@{}}1.1\\[-2pt]\swp{$[0.9,\,1.4]$}\end{tabular}
  & 91\% \\

  & 2024.8
  & \begin{tabular}[c]{@{}c@{}}2.3\\[-2pt]\swp{$[0.9,\,12]$}\end{tabular}
  & \begin{tabular}[c]{@{}c@{}}284\\[-2pt]\swp{$[50,\,359]$}\end{tabular}
  & \begin{tabular}[c]{@{}c@{}}45\\[-2pt]\swp{$[8,\,57]$}\end{tabular}
  & \begin{tabular}[c]{@{}c@{}}0.34\\[-2pt]\swp{$[0.13,\,1.7]$}\end{tabular}
  & \begin{tabular}[c]{@{}c@{}}1.53\\[-2pt]\swp{$[1.35,\,1.93]$}\end{tabular}
  & \begin{tabular}[c]{@{}c@{}}1.4\\[-2pt]\swp{$[1.0,\,2.7]$}\end{tabular}
  & 96\% \\
  \noalign{\smallskip}

\multirow{2}{*}{\vspace{-6pt}YZ Cet c}
  & 2023.9
  & \begin{tabular}[c]{@{}c@{}}1.2\\[-2pt]\swp{$[0.6,\,3.5]$}\end{tabular}
  & \begin{tabular}[c]{@{}c@{}}257\\[-2pt]\swp{$[147,\,332]$}\end{tabular}
  & \begin{tabular}[c]{@{}c@{}}14\\[-2pt]\swp{$[2,\,17]$}\end{tabular}
  & \begin{tabular}[c]{@{}c@{}}0.40\\[-2pt]\swp{$[0.15,\,1.1]$}\end{tabular}
  & \begin{tabular}[c]{@{}c@{}}1.39\\[-2pt]\swp{$[1.22,\,1.61]$}\end{tabular}
  & \begin{tabular}[c]{@{}c@{}}0.7\\[-2pt]\swp{$[0.5,\,1.1]$}\end{tabular}
  & 84\% \\

  & 2024.8
  & \begin{tabular}[c]{@{}c@{}}1.0\\[-2pt]\swp{$[0.4,\,5.7]$}\end{tabular}
  & \begin{tabular}[c]{@{}c@{}}323\\[-2pt]\swp{$[117,\,390]$}\end{tabular}
  & \begin{tabular}[c]{@{}c@{}}25\\[-2pt]\swp{$[4,\,30]$}\end{tabular}
  & \begin{tabular}[c]{@{}c@{}}0.26\\[-2pt]\swp{$[0.14,\,1.2]$}\end{tabular}
  & \begin{tabular}[c]{@{}c@{}}1.45\\[-2pt]\swp{$[1.31,\,1.78]$}\end{tabular}
  & \begin{tabular}[c]{@{}c@{}}0.8\\[-2pt]\swp{$[0.6,\,1.6]$}\end{tabular}
  & 94\% \\
  \noalign{\smallskip}

\multirow{2}{*}{\vspace{-6pt}YZ Cet d}
  & 2023.9
  & \begin{tabular}[c]{@{}c@{}}0.6\\[-2pt]\swp{$[0.3,\,1.7]$}\end{tabular}
  & \begin{tabular}[c]{@{}c@{}}294\\[-2pt]\swp{$[212,\,361]$}\end{tabular}
  & \begin{tabular}[c]{@{}c@{}}8\\[-2pt]\swp{$[0.4,\,9.7]$}\end{tabular}
  & \begin{tabular}[c]{@{}c@{}}0.30\\[-2pt]\swp{$[0.15,\,0.73]$}\end{tabular}
  & \begin{tabular}[c]{@{}c@{}}1.31\\[-2pt]\swp{$[1.16,\,1.49]$}\end{tabular}
  & \begin{tabular}[c]{@{}c@{}}0.5\\[-2pt]\swp{$[0.3,\,0.9]$}\end{tabular}
  & 74\% \\

  & 2024.8
  & \begin{tabular}[c]{@{}c@{}}0.5\\[-2pt]\swp{$[0.2,\,2.7]$}\end{tabular}
  & \begin{tabular}[c]{@{}c@{}}357\\[-2pt]\swp{$[182,\,419]$}\end{tabular}
  & \begin{tabular}[c]{@{}c@{}}14\\[-2pt]\swp{$[1.6,\,16]$}\end{tabular}
  & \begin{tabular}[c]{@{}c@{}}0.24\\[-2pt]\swp{$[0.15,\,0.82]$}\end{tabular}
  & \begin{tabular}[c]{@{}c@{}}1.38\\[-2pt]\swp{$[1.26,\,1.64]$}\end{tabular}
  & \begin{tabular}[c]{@{}c@{}}0.5\\[-2pt]\swp{$[0.3,\,1.2]$}\end{tabular}
  & 91\% \\
  \noalign{\smallskip}

\multirow{3}{*}{\vspace{-18pt}GJ 1151 c}
  & 2020.4
  & \begin{tabular}[c]{@{}c@{}}$1.8 \times 10^{-4}$\\[-2pt]\swp{$[1.6,\,2.0]\times10^{-4}$}\end{tabular}
  & \begin{tabular}[c]{@{}c@{}}366\\[-2pt]\swp{$[360,\,373]$}\end{tabular}
  & \begin{tabular}[c]{@{}c@{}}$2 \times 10^{-3}$\\[-2pt]\swp{$[1.3,\,25]\times10^{-4}$}\end{tabular}
  & \begin{tabular}[c]{@{}c@{}}\swp{$4.6\times10^{-3}$}\\[-2pt]\swp{$[4.2,\,5.2]\times10^{-3}$}\end{tabular}
  & \begin{tabular}[c]{@{}c@{}}\swp{$8.2 \times 10^{-2}$}\\[-2pt]\swp{$[7.7,\,8.4]\times 10^{-2}$}\end{tabular}
  & \begin{tabular}[c]{@{}c@{}}$2.4 \times 10^{-4}$\\[-2pt]\swp{$[2.2,\,2.7]\times10^{-4}$}\end{tabular}
  & 0\% \\

  & 2021.2
  & \begin{tabular}[c]{@{}c@{}}$1.3 \times 10^{-4}$\\[-2pt]\swpt{$[0.6,\,3.5]\times10^{-4}$}\end{tabular}
  & \begin{tabular}[c]{@{}c@{}}458\\[-2pt]\swp{$[351,\,527]$}\end{tabular}
  & \begin{tabular}[c]{@{}c@{}}$7 \times 10^{-3}$\\[-2pt]\swp{$[1.7,\,8.3]\times10^{-3}$}\end{tabular}
  & \begin{tabular}[c]{@{}c@{}}\swp{$1.6\times10^{-2}$}\\[-2pt]\swp{$[0.4,\,2.6]\times10^{-2}$}\end{tabular}
  & \begin{tabular}[c]{@{}c@{}}\swpt{$8.0\times 10^{-2}$}\\[-2pt]\swp{$[7.4,\,8.8]\times 10^{-2}$}\end{tabular}
  & \begin{tabular}[c]{@{}c@{}}$2.3 \times 10^{-4}$\\[-2pt]\swp{$[1.6,\,4.4]\times10^{-4}$}\end{tabular}
  & 0\% \\

  & 2022.2
  & \begin{tabular}[c]{@{}c@{}}$2.8 \times 10^{-4}$\\[-2pt]\swp{$[1.4,\,10]\times10^{-4}$}\end{tabular}
  & \begin{tabular}[c]{@{}c@{}}438\\[-2pt]\swp{$[339,\,483]$}\end{tabular}
  & \begin{tabular}[c]{@{}c@{}}$13 \times 10^{-3}$\\[-2pt]\swp{$[1.4,\,15]\times10^{-3}$}\end{tabular}
  & \begin{tabular}[c]{@{}c@{}}\swp{$1.5 \times 10^{-2}$}\\[-2pt]\swp{$[0.5,\,2.1]\times 10^{-2}$}\end{tabular}
  & \begin{tabular}[c]{@{}c@{}}\swp{$7.4\times10^{-2}$}\\[-2pt]\swp{$[6.8,\,8.0]\times 10^{-2}$}\end{tabular}
  & \begin{tabular}[c]{@{}c@{}}$4.7 \times 10^{-4}$\\[-2pt]\swp{$[3.2,\,12]\times10^{-4}$}\end{tabular}
  & 0\% \\ \hline

\end{tabular}}
\vspace{10pt}
\label{tab:sw_at_orbits_div10}
\end{table*}

\subsection{Comparison with observations}
M dwarfs have been a target of multiple studies attempting to quantify properties of stellar wind. In particular, mass-loss rate can be constrained using both numerical methods and observations. Here we provide a comparison of our sample to observationally derived constraints on mass-loss rate. \swp{We do not make a comparison to polytropic wind models \citep[e.g.][]{Vidotto2014b, See2017}.}

\begin{figure}[ht!]
    \centering
    \includegraphics[width=1\linewidth]{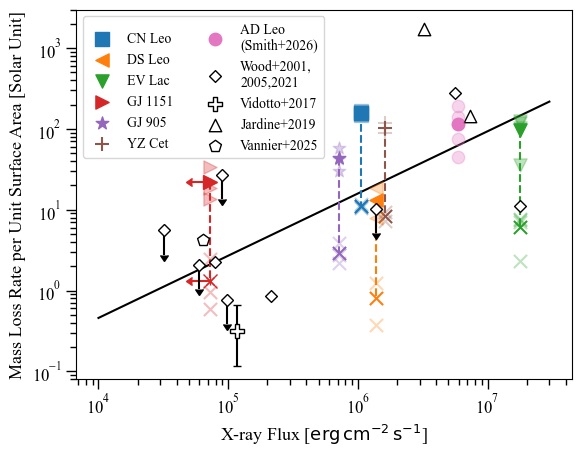}
    \caption{Mass-loss rate per unit surface area versus X-ray flux. Colored points indicate stars for which multiple epochs were simulated using the \swpt{\texttt{AWSoM}} model - star markers indicate output of simulations with $S_A/B = 1.1\times10^5~{\rm erg~cm^{-2}~s^{-1}~G^{-1}}$ while x markers correspond to simulations with $S_A/B = 1.1\times10^4~{\rm erg~cm^{-2}~s^{-1}~G^{-1}}$. The solid points represent the average between epochs, while the translucent points indicate outputs from individual ZDI maps. The value of X-ray fluxes likely vary between epochs as well, however, we do not have contemporaneous measurements with magnetic field maps and thus assume literature values of X-ray flux. White points with black outline indicate observational estimates of M dwarfs' mass-loss rates from literature. The solid line represents the $F_X^{0.77}$ fit from \cite{Wood2021}.}
    \label{fig:mdot_vs_xray}
\end{figure}

In Fig. \ref{fig:mdot_vs_xray} we plotted the mass-loss rate per unit surface area against X-ray flux. The M dwarfs in our sample occupy a parameter space similar to the sample presented in the literature. However, our stars are on the higher end of the range of X-ray fluxes with the exception of GJ 1151. We also provided measurements from other studies, and plotted the scaling law from \cite{Wood2021} as a black line. We see that our results are consistent with the scaling law. With Solar value of $S_A/B$ we found that GJ 1151, \swp{GJ} 905, CN Leo, and YZ Cet have a higher mass-loss rate than that predicted by the scaling law. While for the reduced value of $S_A/B$ we found all mass loss rates per unit surface area in our sample to be at or below the values predicted by the scaling law.

\cite{Vidotto2017} obtained mass-loss rate for GJ 436, an M2.5-type star, by observing atmospheric escape of an exoplanet in orbit around the star. \cite{Jardine2019} calculated $\dot{M}$ of two M dwarfs, V374 Peg and HK Aqr, by observing absorption features in slingshot prominences. \swp{\cite{Wood2001, Wood2005, Wood2021}} derived $\dot{M}$ (or upper bounds in case of a non-detection) for a sample of M dwarfs \swp{(Proxima Cen, \swpt{Barnard's Star,} EV Lac, GJ 411, GJ 887, GJ 273, GJ 205, GJ 588, and YZ CMi)} based on Ly$\alpha$ absorption from interaction of the stellar wind with the interstellar medium. Their M dwarfs span spectral types between \swp{M1.5} and M5.5, radii between 0.14 and 0.59 $R_\odot$, and log X-ray luminosities between \swpt{$10^{25.85}~{\rm erg~s^{-1}}$ and $10^{28.99}~{\rm erg~s^{-1}}$}. \swp{Similar methodology was used by \cite{Vannier2025} to obtain the mass-loss rate for GJ 173. \cite{Wood2021}} proposed the mass-loss rate per unit surface area scales with X-ray flux, $(\dot{M}/A) \propto F_X^{0.77\pm0.04}$. The scatter of the average of each star is consistent with the scatter of data points from the literature. The differences between individual epochs is significant, but it is smaller than the deviation of literature data points away from the scaling law. Thus, the scatter of mass loss rates found in the literature is likely not caused by temporal variability and differences in X-ray flux alone.

Both \swp{\cite{Wood2005}} and our work obtained the mass-loss rate of EV Lac. They report $2.0 \times10^{-14}~M_\odot /yr$, while our calculations show values between $6.4 \times10^{-14}~M_\odot /yr$ and $2.3 \times10^{-13}~M_\odot /yr$. When the Poynting flux-to-magnetic field ratio is decreased by 10 (see Sect. \ref{section:poynting}), our calculations yield $\dot{M}$ between $4.1 \times10^{-15}~M_\odot /yr$ and $1.4 \times10^{-14}~M_\odot /yr$ showing a better agreement with the values presented by \swp{\cite{Wood2005}}. The fact that choosing a lower Poynting\swpt{-}flux-to-magnetic\swpt{-}field ratio than the Solar value is a more suitable option for M dwarfs is in agreement with \cite{Vidotto2023}. As the Ly $\alpha$ observations of \swp{\cite{Wood2005}} and the ZDI maps analyzed here were not simultaneous, the discrepancy between the value may be due to temporal variability of the  magnetic field. 

\section{Conclusions}\label{section:conclusions}
In this paper, we have studied the temporal variations of the magnetically-driven stellar wind and space environment around a sample of six M dwarfs. These M dwarfs have masses between $0.1~M_\odot$ and $0.58~M_\odot$, rotation periods between 2.7 and 175.6 days, and spectral type ranging from M1.0 to M5.5.

Based on the reconstructed ZDI maps of the large-scale magnetic field, we simulated the stellar wind and environment of the six M dwarfs. We computed the mass-loss rate, the angular momentum-loss rate, and the average Alfvén radius for each simulation output. We obtained mass-loss rates ranging between $1.0\times10^{-14}~M_\odot/\rm yr$ and \swp{$2.3\times 10^{-13}~M_\odot/\rm yr$}, angular momentum-loss rates between \swp{$4.6\times10^{27}~\rm erg$} and \swp{$4.9\times10^{31}~\rm erg$}, and average Alfvén radii between \swpt{$16~R_\star$} and \swpt{$41~R_\star$}. Where possible, we made a comparison with observationally constrained mass-loss rates of M dwarfs. Our values are consistent with what is expected of M dwarfs \citep[][]{Wood2021, Vidotto2023}.

For the stars in our sample that have confirmed exoplanets (\swp{DS Leo}, GJ 1151, and YZ Cet), we analyzed the stellar wind conditions at the planetary orbits, including potential magnetic star-planet interactions. Due to the changes of stellar wind properties, the Alfvénic regime at exoplanet orbits can change. \swp{For example, for YZ Cet c, we found that the  fraction of  the orbit within the sub-Alfvénic regime changed from 22\% to 83\%.} In particular, we found that the planet YZ Cet d was completely super-Alfvénic in the 2023.9 epoch and then just a year later in the 2024.8 epoch 45\% of the orbit was super-Alfvénic. These temporal variations change the nature of star-planet interactions
\citep[see e.g.][]{Vidotto2025}.

We performed simulations with both Solar value of Poynting-flux-to-magnetic-field ratio $S_A/B = 1.1\times10^5~{\rm erg~cm^{-2}~s^{-1}~G^{-1}}$ and a reduced value of $S_A/B = 1.1\times10^4~{\rm erg~cm^{-2}~s^{-1}~G^{-1}}$. We found that $\dot{M}$ and $\dot{J}$ differ between these two simulation sets by a constant factor, with both of these quantities increasing with $S_A/B$, and thus this free parameter does not affect our conclusions on the temporal variations of mass- and angular momentum-loss rates. The average Alfvén radius has a non-trivial dependence on $S_A/B$, but we found that it consistently increases as a result of lowering $S_A/B$.

Finally, we compared our values of mass-loss rate against the scaling law of $\dot{M}$ per unit surface area with X-ray flux presented in \cite{Wood2021}. We found our results to be in agreement with the scaling law. Simulations using Solar $S_A/B$ generally yield mass-loss rates higher than those predicted by the scaling law, while the simulations with Solar $S_A/B$ reduced by 10 fall below the scaling law. The scatter of our data points due to temporal variability is lower than the deviations of literature mass-loss rates away from the scaling law, implying that factors others than temporal variability of large-scale magnetic field and amount of X-ray flux cause the observed scatter of mass-loss rate between stars.

Future studies of stellar magnetic environments around M dwarf stars should incorporate a larger samples of magnetic field maps obtained with ZDI. Given the limited temporal baseline to capture magnetic cycles for the six stars examined in this work, we also emphasize the need of follow-up spectropolarimetric observations in order to pose stronger constraints on magnetic variability. Since such temporal variations of the magnetic field translate to a change of the Alfvén surface, we advocate for quasi-simultaneous radio and spectropolarimetric observations to correctly interpret potential detections of SPI signatures.

\begin{acknowledgements}
       \swpt{We would like to thank the anonymous referee for the constructive comments that improved this manuscript.} SB and AAV acknowledge funding by the Dutch Research Council (NWO) under the project \swp{``}Exo-space weather and contemporaneous signatures of star-planet interactions" (with project number OCENW.M.22.215 of the research programme \swp{``}Open Competition Domain Science- M"). AAV acknowledges funding from the Dutch Research Council (NWO), with project number VI.C.232.041 of the Talent Programme Vici. We acknowledge Python libraries without which the analysis presented in this work would not be possible: \swp{\texttt{NumPy} \citep{NumPy}, \texttt{SciPy} \citep{SciPy}, \texttt{AstroPy} \citep{AstroPy}, \texttt{Matplotlib} \citep{MatPlotLib}, \texttt{Scikit-Image} \citep{scikit-image}.} This work used the Dutch national e-infrastructure with the support of the SURF Cooperative (grant EINF-13632). This work used the BATS-R-US tools developed at the University of Michigan Center for Space Environment Modeling and made available through the NASA Community Coordinated Modeling Center.
\end{acknowledgements}

\bibliographystyle{bibtex/aa}
\bibliography{references}

@ARTICLE{1993Icar..101..108K,
       author = {{Kasting}, James F. and {Whitmire}, Daniel P. and {Reynolds}, Ray T.},
        title = "{Habitable Zones around Main Sequence Stars}",
      journal = {\icarus},
         year = 1993,
        month = jan,
       volume = {101},
       number = {1},
        pages = {108-128},
          doi = {10.1006/icar.1993.1010},
       adsurl = {https://ui.adsabs.harvard.edu/abs/1993Icar..101..108K}
}

@ARTICLE{Kopparapu2013b,
       author = {{Kopparapu}, Ravi Kumar and {Ramirez}, Ramses and {Kasting}, James F. and {Eymet}, Vincent and {Robinson}, Tyler D. and {Mahadevan}, Suvrath and {Terrien}, Ryan C. and {Domagal-Goldman}, Shawn and {Meadows}, Victoria and {Deshpande}, Rohit},
        title = "{Habitable Zones around Main-sequence Stars: New Estimates}",
      journal = {\apj},
         year = 2013,
        month = mar,
       volume = {765},
       number = {2},
          eid = {131},
        pages = {131},
          doi = {10.1088/0004-637X/765/2/131},
archivePrefix = {arXiv},
       eprint = {1301.6674},
 primaryClass = {astro-ph.EP},
       adsurl = {https://ui.adsabs.harvard.edu/abs/2013ApJ...765..131K}
}

@ARTICLE{Revilla2026,
       author = {{Revilla}, D. and {Amado}, P.~J. and {Luque}, R. and {Sch{\"o}fer}, P. and {Lanza}, A.~F. and {Binnenfeld}, A. and {Caballero}, J.~A. and {Hatzes}, A.~P. and {Henry}, G.~W. and {Jeffers}, S.~V. and {Kaur}, S. and {Pall{\'e}}, E. and {Pe{\~n}a-Mo{\~n}ino}, L. and {P{\'e}rez-Torres}, M. and {Quirrenbach}, A. and {Reiners}, A. and {Ribas}, I. and {Vigan{\`o}}, D. and {Zapatero-Osorio}, M.~R. and {Zucker}, S.},
        title = "{Constraining an exoplanet{\textquoteright}s magnetic field using star-planet interactions}",
      journal = {Science},
         year = 2026,
        month = jul,
       volume = {393},
       number = {6809},
        pages = {403-407},
          doi = {10.1126/science.adv3075},
archivePrefix = {arXiv},
       eprint = {2604.06969},
 primaryClass = {astro-ph.EP},
       adsurl = {https://ui.adsabs.harvard.edu/abs/2026Sci...393..403R}
}

@ARTICLE{ZapateroOsorio2026,
       author = {{Zapatero Osorio}, M.~R. and {B{\'e}jar}, V.~J.~S. and {Su{\'a}rez Mascare{\~n}o}, A. and {Rebolo}, R. and {Cristiani}, S. and {Micela}, G. and {Miles-P{\'a}ez}, P.~A. and {Santos}, N.~C. and {Pall{\'e}}, E. and {Gonz{\'a}lez Hern{\'a}ndez}, J.~I. and {Damasso}, M. and {Castro-Gonz{\'a}lez}, A. and {Martins}, C.~J.~A.~P. and {Pepe}, F. and {Sozzetti}, A. and {Lavie}, B. and {Rodrigues}, J.},
        title = "{Star─planet interaction in the Proxima system}",
      journal = {\aap},
         year = 2026,
        month = jul,
       volume = {711},
          eid = {A116},
        pages = {A116},
          doi = {10.1051/0004-6361/202660158},
archivePrefix = {arXiv},
       eprint = {2605.22925},
 primaryClass = {astro-ph.EP},
       adsurl = {https://ui.adsabs.harvard.edu/abs/2026A&A...711A.116Z}
}

@ARTICLE{Shkolnik2005,
       author = {{Shkolnik}, E. and {Walker}, G.~A.~H. and {Bohlender}, D.~A. and {Gu}, P.-G. and {K{\"u}rster}, M.},
        title = "{Hot Jupiters and Hot Spots: The Short- and Long-Term Chromospheric Activity on Stars with Giant Planets}",
      journal = {\apj},
         year = 2005,
        month = apr,
       volume = {622},
       number = {2},
        pages = {1075-1090},
          doi = {10.1086/428037},
archivePrefix = {arXiv},
       eprint = {astro-ph/0411655},
 primaryClass = {astro-ph},
       adsurl = {https://ui.adsabs.harvard.edu/abs/2005ApJ...622.1075S}
}

@ARTICLE{Shkolnik2003,
       author = {{Shkolnik}, E. and {Walker}, G.~A.~H. and {Bohlender}, D.~A.},
        title = "{Evidence for Planet-induced Chromospheric Activity on HD 179949}",
      journal = {\apj},
         year = 2003,
        month = nov,
       volume = {597},
       number = {2},
        pages = {1092-1096},
          doi = {10.1086/378583},
       adsurl = {https://ui.adsabs.harvard.edu/abs/2003ApJ...597.1092S}
}

@ARTICLE{Vannier2025,
       author = {{Vannier}, Hunter and {Redfield}, Seth and {Wood}, Brian E. and {M{\"u}ller}, Hans-Reinhard and {Linsky}, Jeffrey L. and {Frisch}, Priscilla C.},
        title = "{Mapping Our Path through the Local Interstellar Medium: High-resolution Ultraviolet Absorption Spectroscopy of Sight Lines along the Sun's Historical Trajectory}",
      journal = {\apj},
         year = 2025,
        month = mar,
       volume = {981},
       number = {2},
          eid = {102},
        pages = {102},
          doi = {10.3847/1538-4357/adb033},
       adsurl = {https://ui.adsabs.harvard.edu/abs/2025ApJ...981..102V}
}

@ARTICLE{Wood2005,
       author = {{Wood}, Brian E. and {Redfield}, Seth and {Linsky}, Jeffrey L. and {M{\"u}ller}, Hans-Reinhard and {Zank}, Gary P.},
        title = "{Stellar Ly{\ensuremath{\alpha}} Emission Lines in the Hubble Space Telescope Archive: Intrinsic Line Fluxes and Absorption from the Heliosphere and Astrospheres}",
      journal = {\apjs},
         year = 2005,
        month = jul,
       volume = {159},
       number = {1},
        pages = {118-140},
          doi = {10.1086/430523},
archivePrefix = {arXiv},
       eprint = {astro-ph/0503372},
 primaryClass = {astro-ph},
       adsurl = {https://ui.adsabs.harvard.edu/abs/2005ApJS..159..118W}
}

@ARTICLE{Wood2001,
       author = {{Wood}, Brian E. and {Linsky}, Jeffrey L. and {M{\"u}ller}, Hans-Reinhard and {Zank}, Gary P.},
        title = "{Observational Estimates for the Mass-Loss Rates of {\ensuremath{\alpha}} Centauri and Proxima Centauri Using Hubble Space Telescope Ly{\ensuremath{\alpha}} Spectra}",
      journal = {\apjl},
         year = 2001,
        month = jan,
       volume = {547},
       number = {1},
        pages = {L49-L52},
          doi = {10.1086/318888},
archivePrefix = {arXiv},
       eprint = {astro-ph/0011153},
 primaryClass = {astro-ph},
       adsurl = {https://ui.adsabs.harvard.edu/abs/2001ApJ...547L..49W}
}

@ARTICLE{Scandariato2013,
       author = {{Scandariato}, G. and {Maggio}, A. and {Lanza}, A.~F. and {Pagano}, I. and {Fares}, R. and {Shkolnik}, E.~L. and {Bohlender}, D. and {Cameron}, A.~C. and {Dieters}, S. and {Donati}, J.-F. and {Mart{\'\i}nez Fiorenzano}, A.~F. and {Jardine}, M. and {Moutou}, C.},
        title = "{A coordinated optical and X-ray spectroscopic campaign on HD 179949: searching for planet-induced chromospheric and coronal activity}",
      journal = {\aap},
         year = 2013,
        month = apr,
       volume = {552},
          eid = {A7},
        pages = {A7},
          doi = {10.1051/0004-6361/201219875},
archivePrefix = {arXiv},
       eprint = {1301.7748},
 primaryClass = {astro-ph.SR},
       adsurl = {https://ui.adsabs.harvard.edu/abs/2013A&A...552A...7S}
}

@ARTICLE{Donati2025,
       author = {{Donati}, J.-F. and {Cristofari}, P.~I. and {Klein}, B. and {Finociety}, B. and {Moutou}, C.},
        title = "{Full Stokes magnetometry of the active M dwarfs AU Mic and EV Lac with SPIRou}",
      journal = {\aap},
         year = 2025,
        month = aug,
       volume = {700},
          eid = {A122},
        pages = {A122},
          doi = {10.1051/0004-6361/202555428},
archivePrefix = {arXiv},
       eprint = {2507.01754},
 primaryClass = {astro-ph.SR},
       adsurl = {https://ui.adsabs.harvard.edu/abs/2025A&A...700A.122D}
}

@ARTICLE{Blackman2018,
       author = {{Blackman}, Eric G. and {Tarduno}, John A.},
        title = "{Mass, energy, and momentum capture from stellar winds by magnetized and unmagnetized planets: implications for atmospheric erosion and habitability}",
      journal = {\mnras},
         year = 2018,
        month = dec,
       volume = {481},
       number = {4},
        pages = {5146-5155},
          doi = {10.1093/mnras/sty2640},
archivePrefix = {arXiv},
       eprint = {1801.00895},
 primaryClass = {astro-ph.EP},
       adsurl = {https://ui.adsabs.harvard.edu/abs/2018MNRAS.481.5146B}
}

@Article{Fionnagain2021,
  author        = {{O Fionnagain}, D. and {Vidotto}, A.~A. and {Petit}, P. and {Neiner}, C. and {Manchester}, W., IV and {Folsom}, C.~P. and {Hallinan}, G.},
  journal       = {\mnras},
  title         = {{lambda And: a post-main-sequence wind from a solar-mass star}},
  year          = {2021},
  month         = jan,
  number        = {3},
  pages         = {3438-3453},
  volume        = {500},
  adsurl        = {https://ui.adsabs.harvard.edu/abs/2021MNRAS.500.3438O},
  archiveprefix = {arXiv},
  doi           = {10.1093/mnras/staa3468},
  eprint        = {2011.02406},
  primaryclass  = {astro-ph.SR},
}

@ARTICLE{Presa2026,
       author = {{Presa}, Andr{\'e}s and {Vidotto}, Aline A. and {Elekes}, Filip},
        title = "{Star─planet magnetic interactions in photoevaporating exoplanets: enhanced power due to atmospheric escape}",
      journal = {\mnras},
         year = 2026,
        month = jul,
       volume = {549},
       number = {4},
          eid = {stag1081},
        pages = {stag1081},
          doi = {10.1093/mnras/stag1081},
archivePrefix = {arXiv},
       eprint = {2604.06064},
 primaryClass = {astro-ph.EP},
       adsurl = {https://ui.adsabs.harvard.edu/abs/2026MNRAS.549g1081P}
}

@ARTICLE{Schmitt2004,
       author = {{Schmitt}, J.~H.~M.~M. and {Liefke}, C.},
        title = "{NEXXUS: A comprehensive ROSAT survey of coronal X-ray emission among nearby solar-like stars}",
      journal = {\aap},
         year = 2004,
        month = apr,
       volume = {417},
        pages = {651-665},
          doi = {10.1051/0004-6361:20030495},
archivePrefix = {arXiv},
       eprint = {astro-ph/0308510},
 primaryClass = {astro-ph},
       adsurl = {https://ui.adsabs.harvard.edu/abs/2004A&A...417..651S}
}

@ARTICLE{Carolan2020,
       author = {{Carolan}, S. and {Vidotto}, A.~A. and {Plavchan}, P. and {Villarreal D'Angelo}, C. and {Hazra}, G.},
        title = "{The dichotomy of atmospheric escape in AU Mic b}",
      journal = {\mnras},
         year = 2020,
        month = nov,
       volume = {498},
       number = {1},
        pages = {L53-L57},
          doi = {10.1093/mnrasl/slaa127},
archivePrefix = {arXiv},
       eprint = {2006.13606},
 primaryClass = {astro-ph.EP},
       adsurl = {https://ui.adsabs.harvard.edu/abs/2020MNRAS.498L..53C}
}

@ARTICLE{Carolan2019,
       author = {{Carolan}, S. and {Vidotto}, A.~A. and {Loesch}, C. and {Coogan}, P.},
        title = "{The evolution of Earth's magnetosphere during the solar main sequence}",
      journal = {\mnras},
         year = 2019,
        month = nov,
       volume = {489},
       number = {4},
        pages = {5784-5801},
          doi = {10.1093/mnras/stz2422},
archivePrefix = {arXiv},
       eprint = {1908.03537},
 primaryClass = {astro-ph.EP},
       adsurl = {https://ui.adsabs.harvard.edu/abs/2019MNRAS.489.5784C}
}

@ARTICLE{Slavin2014,
       author = {{Slavin}, James A. and {DiBraccio}, Gina A. and {Gershman}, Daniel J. and {Imber}, Suzanne M. and {Poh}, Gang Kai and {Raines}, Jim M. and {Zurbuchen}, Thomas H. and {Jia}, Xianzhe and {Baker}, Daniel N. and {Glassmeier}, Karl-Heinz and {Livi}, Stefano A. and {Boardsen}, Scott A. and {Cassidy}, Timothy A. and {Sarantos}, Menelaos and {Sundberg}, Torbjorn and {Masters}, Adam and {Johnson}, Catherine L. and {Winslow}, Reka M. and {Anderson}, Brian J. and {Korth}, Haje and {McNutt}, Ralph L. and {Solomon}, Sean C.},
        title = "{MESSENGER observations of Mercury's dayside magnetosphere under extreme solar wind conditions}",
      journal = {Journal of Geophysical Research (Space Physics)},
         year = 2014,
        month = oct,
       volume = {119},
       number = {10},
        pages = {8087-8116},
          doi = {10.1002/2014JA020319},
       adsurl = {https://ui.adsabs.harvard.edu/abs/2014JGRA..119.8087S}
}

@INCOLLECTION{Bagenal2013,
       author = {{Bagenal}, Fran},
        title = "{Planetary Magnetospheres}",
    booktitle = {Planets, Stars and Stellar Systems. Volume 3: Solar and Stellar Planetary Systems},
         year = 2013,
       editor = {{Oswalt}, Terry D. and {French}, Linda M. and {Kalas}, Paul},
        pages = {251},
          doi = {10.1007/978-94-007-5606-9_6},
       adsurl = {https://ui.adsabs.harvard.edu/abs/2013pss3.book..251B}
}

@ARTICLE{Kavanagh2021,
       author = {{Kavanagh}, Robert D. and {Vidotto}, Aline A. and {Klein}, Baptiste and {Jardine}, Moira M. and {Donati}, Jean-Fran{\c{c}}ois and {{\'O} Fionnag{\'a}in}, D{\'u}alta},
        title = "{Planet-induced radio emission from the coronae of M dwarfs: the case of Prox Cen and AU Mic}",
      journal = {\mnras},
         year = 2021,
        month = jun,
       volume = {504},
       number = {1},
        pages = {1511-1518},
          doi = {10.1093/mnras/stab929},
archivePrefix = {arXiv},
       eprint = {2103.16318},
 primaryClass = {astro-ph.SR},
       adsurl = {https://ui.adsabs.harvard.edu/abs/2021MNRAS.504.1511K}
}

@ARTICLE{Presa2024,
       author = {{Presa}, Andr{\'e}s and {Driessen}, Florian A. and {Vidotto}, Aline A.},
        title = "{Atmospheric escape in hot Jupiters under sub-Alfv{\'e}nic interactions}",
      journal = {\mnras},
         year = 2024,
        month = nov,
       volume = {534},
       number = {4},
        pages = {3622-3636},
          doi = {10.1093/mnras/stae2325},
archivePrefix = {arXiv},
       eprint = {2410.08881},
 primaryClass = {astro-ph.EP},
       adsurl = {https://ui.adsabs.harvard.edu/abs/2024MNRAS.534.3622P}
}

@ARTICLE{Zeeman1897,
       author = {{Zeeman}, P.},
        title = "{On the Influence of Magnetism on the Nature of the Light Emitted by a Substance.}",
      journal = {\apj},
         year = 1897,
        month = may,
       volume = {5},
        pages = {332},
          doi = {10.1086/140355},
       adsurl = {https://ui.adsabs.harvard.edu/abs/1897ApJ.....5..332Z}
}

@ARTICLE{Jivani2023,
       author = {{Jivani}, Aniket and {Sachdeva}, Nishtha and {Huang}, Zhenguang and {Chen}, Yang and {van der Holst}, Bart and {Manchester}, Ward and {Iong}, Daniel and {Chen}, Hongfan and {Zou}, Shasha and {Huan}, Xun and {Toth}, Gabor},
        title = "{Global Sensitivity Analysis and Uncertainty Quantification for Background Solar Wind Using the Alfv{\'e}n Wave Solar Atmosphere Model}",
      journal = {Space Weather},
         year = 2023,
        month = jan,
       volume = {21},
       number = {1},
          eid = {e2022SW003262},
        pages = {e2022SW003262},
          doi = {10.1029/2022SW003262},
       adsurl = {https://ui.adsabs.harvard.edu/abs/2023SpWea..2103262J}
}

@ARTICLE{Sachdeva2019,
       author = {{Sachdeva}, Nishtha and {van der Holst}, Bart and {Manchester}, Ward B. and {T{\'o}th}, Gabor and {Chen}, Yuxi and {Lloveras}, Diego G. and {V{\'a}squez}, Alberto M. and {Lamy}, Philippe and {Wojak}, Julien and {Jackson}, Bernard V. and {Yu}, Hsiu-Shan and {Henney}, Carl J.},
        title = "{Validation of the Alfv{\'e}n Wave Solar Atmosphere Model (AWSoM) with Observations from the Low Corona to 1 au}",
      journal = {\apj},
         year = 2019,
        month = dec,
       volume = {887},
       number = {1},
          eid = {83},
        pages = {83},
          doi = {10.3847/1538-4357/ab4f5e},
archivePrefix = {arXiv},
       eprint = {1910.08110},
 primaryClass = {astro-ph.SR},
       adsurl = {https://ui.adsabs.harvard.edu/abs/2019ApJ...887...83S}
}

@ARTICLE{vanderHolst2019,
       author = {{van der Holst}, B. and {Manchester}, IV, W.~B. and {Klein}, K.~G. and {Kasper}, J.~C.},
        title = "{Predictions for the First Parker Solar Probe Encounter}",
      journal = {\apjl},
         year = 2019,
        month = feb,
       volume = {872},
       number = {2},
          eid = {L18},
        pages = {L18},
          doi = {10.3847/2041-8213/ab04a5},
archivePrefix = {arXiv},
       eprint = {1902.03921},
 primaryClass = {physics.space-ph},
       adsurl = {https://ui.adsabs.harvard.edu/abs/2019ApJ...872L..18V}
}

@ARTICLE{Meng2015,
       author = {{Meng}, X. and {van der Holst}, B. and {T{\'o}th}, G. and {Gombosi}, T.~I.},
        title = "{Alfv{\'e}n wave solar model (AWSoM): proton temperature anisotropy and solar wind acceleration}",
      journal = {\mnras},
         year = 2015,
        month = dec,
       volume = {454},
       number = {4},
        pages = {3697-3709},
          doi = {10.1093/mnras/stv2249},
       adsurl = {https://ui.adsabs.harvard.edu/abs/2015MNRAS.454.3697M}
}

@ARTICLE{Trigilio2023,
       author = {{Trigilio}, Corrado and {Biswas}, Ayan and {Leto}, Paolo and {Umana}, Grazia and {Busa}, Innocenza and {Cavallaro}, Francesco and {Das}, Barnali and {Chandra}, Poonam and {Perez-Torres}, Miguel and {Wade}, Gregg A. and {Bordiu}, Cristobal and {Buemi}, Carla S. and {Bufano}, Filomena and {Ingallinera}, Adriano and {Loru}, Sara and {Riggi}, Simone},
        title = "{Star-Planet Interaction at radio wavelengths in YZ Ceti: Inferring planetary magnetic field}",
      journal = {submitted to \apj},
         year = 2023,
        month = may,
       adsurl = {https://ui.adsabs.harvard.edu/abs/2023arXiv230500809T}
}

@ARTICLE{Pineda2026,
       author = {{Pineda}, J. Sebastian and {Bellotti}, Stefano and {Villadsen}, Jackie and {Vidotto}, Aline and {Morin}, Julien and {Folsom}, Colin P.},
        title = "{Evaluating Star─Planet Interactions with Zeeman Doppler Imaging: Case Study in YZ Ceti}",
      journal = {\apj},
         year = 2026,
        month = feb,
       volume = {997},
       number = {2},
          eid = {172},
        pages = {172},
          doi = {10.3847/1538-4357/ae2619},
archivePrefix = {arXiv},
       eprint = {2511.21853},
 primaryClass = {astro-ph.SR},
       adsurl = {https://ui.adsabs.harvard.edu/abs/2026ApJ...997..172P}
}

@ARTICLE{Bourrier2020,
       author = {{Bourrier}, V. and {Wheatley}, P.~J. and {Lecavelier des Etangs}, A. and {King}, G. and {Louden}, T. and {Ehrenreich}, D. and {Fares}, R. and {Helling}, Ch and {Llama}, J. and {Jardine}, M.~M. and {Vidotto}, A.~A.},
        title = "{MOVES III. Simultaneous X-ray and ultraviolet observations unveiling the variable environment of the hot Jupiter HD 189733b}",
      journal = {\mnras},
         year = 2020,
        month = mar,
       volume = {493},
       number = {1},
        pages = {559-579},
          doi = {10.1093/mnras/staa256},
archivePrefix = {arXiv},
       eprint = {2001.11048},
 primaryClass = {astro-ph.EP},
       adsurl = {https://ui.adsabs.harvard.edu/abs/2020MNRAS.493..559B}
}

@ARTICLE{Cauley2018,
       author = {{Cauley}, P. Wilson and {Shkolnik}, Evgenya L. and {Llama}, Joe and {Bourrier}, Vincent and {Moutou}, Claire},
        title = "{Evidence of Magnetic Star-Planet Interactions in the HD 189733 System from Orbitally Phased Ca II K Variations}",
      journal = {\aj},
         year = 2018,
        month = dec,
       volume = {156},
       number = {6},
          eid = {262},
        pages = {262},
          doi = {10.3847/1538-3881/aae841},
archivePrefix = {arXiv},
       eprint = {1810.05253},
 primaryClass = {astro-ph.EP},
       adsurl = {https://ui.adsabs.harvard.edu/abs/2018AJ....156..262C}
}

@ARTICLE{Shkolnik2008,
       author = {{Shkolnik}, Evgenya and {Bohlender}, David A. and {Walker}, Gordon A.~H. and {Collier Cameron}, Andrew},
        title = "{The On/Off Nature of Star-Planet Interactions}",
      journal = {\apj},
         year = 2008,
        month = mar,
       volume = {676},
       number = {1},
        pages = {628-638},
          doi = {10.1086/527351},
archivePrefix = {arXiv},
       eprint = {0712.0004},
 primaryClass = {astro-ph},
       adsurl = {https://ui.adsabs.harvard.edu/abs/2008ApJ...676..628S}
}

@ARTICLE{Cristofari2025,
       author = {{Cristofari}, P.~I. and {Donati}, J.-F. and {Bellotti}, S. and {Artigau}, {\'E}. and {Carmona}, A. and {Moutou}, C. and {Delfosse}, X. and {Petit}, P. and {Finociety}, B. and {Dias do Nascimento}, J.},
        title = "{Rotational modulation and long-term evolution of the small-scale magnetic fields of M dwarfs observed with SPIRou}",
      journal = {\aap},
         year = 2025,
        month = oct,
       volume = {702},
          eid = {A111},
        pages = {A111},
          doi = {10.1051/0004-6361/202554902},
archivePrefix = {arXiv},
       eprint = {2508.04569},
 primaryClass = {astro-ph.SR},
       adsurl = {https://ui.adsabs.harvard.edu/abs/2025A&A...702A.111C}
}

@ARTICLE{See2016,
       author = {{See}, V. and {Jardine}, M. and {Vidotto}, A.~A. and {Donati}, J.-F. and {Boro Saikia}, S. and {Bouvier}, J. and {Fares}, R. and {Folsom}, C.~P. and {Gregory}, S.~G. and {Hussain}, G. and {Jeffers}, S.~V. and {Marsden}, S.~C. and {Morin}, J. and {Moutou}, C. and {do Nascimento}, J.~D. and {Petit}, P. and {Waite}, I.~A.},
        title = "{The connection between stellar activity cycles and magnetic field topology}",
      journal = {\mnras},
         year = 2016,
        month = nov,
       volume = {462},
       number = {4},
        pages = {4442-4450},
          doi = {10.1093/mnras/stw2010},
archivePrefix = {arXiv},
       eprint = {1610.03737},
 primaryClass = {astro-ph.SR},
       adsurl = {https://ui.adsabs.harvard.edu/abs/2016MNRAS.462.4442S}
}

@ARTICLE{Marsden2023,
       author = {{Marsden}, S.~C. and {Evensberget}, D. and {Brown}, E.~L. and {Neiner}, C. and {Seach}, J.~M. and {Morin}, J. and {Petit}, P. and {Jeffers}, S.~V. and {Folsom}, C.~P.},
        title = "{The magnetic field and stellar wind of the mature late-F star {\ensuremath{\chi}} Draconis A}",
      journal = {\mnras},
         year = 2023,
        month = jun,
       volume = {522},
       number = {1},
        pages = {792-810},
          doi = {10.1093/mnras/stad925},
       adsurl = {https://ui.adsabs.harvard.edu/abs/2023MNRAS.522..792M}
}

@article{scikit-image,
 title = {scikit-image: image processing in {P}ython},
 author = {van der Walt, {S}t\'efan and {S}ch\"onberger, {J}ohannes {L}. and
           {Nunez-Iglesias}, {J}uan and {B}oulogne, {F}ran\c{c}ois and {W}arner,
           {J}oshua {D}. and {Y}ager, {N}eil and {G}ouillart, {E}mmanuelle and
           {Y}u, {T}ony and the scikit-image contributors},
 year = {2014},
 month = {6},
 volume = {2},
 pages = {e453},
 journal = {PeerJ},
 issn = {2167-8359},
 url = {https://doi.org/10.7717/peerj.453},
 doi = {10.7717/peerj.453}
}

@Article{MatPlotLib,
  Author    = {Hunter, J. D.},
  Title     = {Matplotlib: A 2D graphics environment},
  Journal   = {Computing in Science \& Engineering},
  Volume    = {9},
  Number    = {3},
  Pages     = {90--95},
  publisher = {IEEE COMPUTER SOC},
  doi       = {10.1109/MCSE.2007.55},
  year      = 2007
}

@ARTICLE{AstroPy,
       author = {{Astropy Collaboration} and {Price-Whelan}, Adrian M. and {Lim}, Pey Lian and {Earl}, Nicholas and {Starkman}, Nathaniel and {Bradley}, Larry and {Shupe}, David L. and {Patil}, Aarya A. and {Corrales}, Lia and {Brasseur}, C.~E. and {N{\"o}the}, Maximilian and {Donath}, Axel and {Tollerud}, Erik and {Morris}, Brett M. and {Ginsburg}, Adam and {Vaher}, Eero and {Weaver}, Benjamin A. and {Tocknell}, James and {Jamieson}, William and {van Kerkwijk}, Marten H. and {Robitaille}, Thomas P. and {Merry}, Bruce and {Bachetti}, Matteo and {G{\"u}nther}, H. Moritz and {Aldcroft}, Thomas L. and {Alvarado-Montes}, Jaime A. and {Archibald}, Anne M. and {B{\'o}di}, Attila and {Bapat}, Shreyas and {Barentsen}, Geert and {Baz{\'a}n}, Juanjo and {Biswas}, Manish and {Boquien}, M{\'e}d{\'e}ric and {Burke}, D.~J. and {Cara}, Daria and {Cara}, Mihai and {Conroy}, Kyle E. and {Conseil}, Simon and {Craig}, Matthew W. and {Cross}, Robert M. and {Cruz}, Kelle L. and {D'Eugenio}, Francesco and {Dencheva}, Nadia and {Devillepoix}, Hadrien A.~R. and {Dietrich}, J{\"o}rg P. and {Eigenbrot}, Arthur Davis and {Erben}, Thomas and {Ferreira}, Leonardo and {Foreman-Mackey}, Daniel and {Fox}, Ryan and {Freij}, Nabil and {Garg}, Suyog and {Geda}, Robel and {Glattly}, Lauren and {Gondhalekar}, Yash and {Gordon}, Karl D. and {Grant}, David and {Greenfield}, Perry and {Groener}, Austen M. and {Guest}, Steve and {Gurovich}, Sebastian and {Handberg}, Rasmus and {Hart}, Akeem and {Hatfield-Dodds}, Zac and {Homeier}, Derek and {Hosseinzadeh}, Griffin and {Jenness}, Tim and {Jones}, Craig K. and {Joseph}, Prajwel and {Kalmbach}, J. Bryce and {Karamehmetoglu}, Emir and {Ka{\l}uszy{\'n}ski}, Miko{\l}aj and {Kelley}, Michael S.~P. and {Kern}, Nicholas and {Kerzendorf}, Wolfgang E. and {Koch}, Eric W. and {Kulumani}, Shankar and {Lee}, Antony and {Ly}, Chun and {Ma}, Zhiyuan and {MacBride}, Conor and {Maljaars}, Jakob M. and {Muna}, Demitri and {Murphy}, N.~A. and {Norman}, Henrik and {O'Steen}, Richard and {Oman}, Kyle A. and {Pacifici}, Camilla and {Pascual}, Sergio and {Pascual-Granado}, J. and {Patil}, Rohit R. and {Perren}, Gabriel I. and {Pickering}, Timothy E. and {Rastogi}, Tanuj and {Roulston}, Benjamin R. and {Ryan}, Daniel F. and {Rykoff}, Eli S. and {Sabater}, Jose and {Sakurikar}, Parikshit and {Salgado}, Jes{\'u}s and {Sanghi}, Aniket and {Saunders}, Nicholas and {Savchenko}, Volodymyr and {Schwardt}, Ludwig and {Seifert-Eckert}, Michael and {Shih}, Albert Y. and {Jain}, Anany Shrey and {Shukla}, Gyanendra and {Sick}, Jonathan and {Simpson}, Chris and {Singanamalla}, Sudheesh and {Singer}, Leo P. and {Singhal}, Jaladh and {Sinha}, Manodeep and {Sip{\H{o}}cz}, Brigitta M. and {Spitler}, Lee R. and {Stansby}, David and {Streicher}, Ole and {{\v{S}}umak}, Jani and {Swinbank}, John D. and {Taranu}, Dan S. and {Tewary}, Nikita and {Tremblay}, Grant R. and {de Val-Borro}, Miguel and {Van Kooten}, Samuel J. and {Vasovi{\'c}}, Zlatan and {Verma}, Shresth and {de Miranda Cardoso}, Jos{\'e} Vin{\'\i}cius and {Williams}, Peter K.~G. and {Wilson}, Tom J. and {Winkel}, Benjamin and {Wood-Vasey}, W.~M. and {Xue}, Rui and {Yoachim}, Peter and {Zhang}, Chen and {Zonca}, Andrea and {Astropy Project Contributors}},
        title = "{The Astropy Project: Sustaining and Growing a Community-oriented Open-source Project and the Latest Major Release (v5.0) of the Core Package}",
      journal = {\apj},
         year = 2022,
        month = aug,
       volume = {935},
       number = {2},
          eid = {167},
        pages = {167},
          doi = {10.3847/1538-4357/ac7c74},
archivePrefix = {arXiv},
       eprint = {2206.14220},
 primaryClass = {astro-ph.IM},
       adsurl = {https://ui.adsabs.harvard.edu/abs/2022ApJ...935..167A}
}

@ARTICLE{Garraffo2016,
       author = {{Garraffo}, C. and {Drake}, J.~J. and {Cohen}, O.},
        title = "{The Space Weather of Proxima Centauri b}",
      journal = {\apjl},
         year = 2016,
        month = dec,
       volume = {833},
       number = {1},
          eid = {L4},
        pages = {L4},
          doi = {10.3847/2041-8205/833/1/L4},
archivePrefix = {arXiv},
       eprint = {1609.09076},
 primaryClass = {astro-ph.SR},
       adsurl = {https://ui.adsabs.harvard.edu/abs/2016ApJ...833L...4G}
}

@ARTICLE{Ilin2025,
       author = {{Ilin}, Ekaterina and {Vedantham}, Harish K. and {Poppenh{\"a}ger}, Katja and {Bloot}, Sanne and {Callingham}, Joseph R. and {Brandeker}, Alexis and {Chakraborty}, Hritam},
        title = "{Close-in planet induces flares on its host star}",
      journal = {\nat},
         year = 2025,
        month = jul,
       volume = {643},
       number = {8072},
        pages = {645-648},
          doi = {10.1038/s41586-025-09236-z},
archivePrefix = {arXiv},
       eprint = {2507.00791},
 primaryClass = {astro-ph.EP},
       adsurl = {https://ui.adsabs.harvard.edu/abs/2025Natur.643..645I}
}

@ARTICLE{Mignon2023,
       author = {{Mignon}, L. and {Meunier}, N. and {Delfosse}, X. and {Bonfils}, X. and {Santos}, N.~C. and {Forveille}, T. and {Gaisn{\'e}}, G. and {Astudillo-Defru}, N. and {Lovis}, C. and {Udry}, S.},
        title = "{Characterisation of stellar activity of M dwarfs. I. Long-timescale variability in a large sample and detection of new cycles}",
      journal = {\aap},
         year = 2023,
        month = jul,
       volume = {675},
          eid = {A168},
        pages = {A168},
          doi = {10.1051/0004-6361/202244249},
archivePrefix = {arXiv},
       eprint = {2303.03998},
 primaryClass = {astro-ph.SR},
       adsurl = {https://ui.adsabs.harvard.edu/abs/2023A&A...675A.168M}
}

@ARTICLE{Oviedo2026,
       author = {{Oviedo}, C.~G. and {Buccino}, A.~P. and {D{\'\i}az}, R.~F. and {Petrucci}, R. and {Jofr{\'e}}, E. and {Boisse}, I. and {Colombo}, P.~D. and {Delfosse}, X.},
        title = "{Long-term activity cycles in planetary M stars observed with SOPHIE}",
      journal = {\aap},
         year = 2026,
        month = mar,
       volume = {708},
          eid = {A107},
        pages = {A107},
          doi = {10.1051/0004-6361/202556087},
archivePrefix = {arXiv},
       eprint = {2603.03643},
 primaryClass = {astro-ph.SR},
       adsurl = {https://ui.adsabs.harvard.edu/abs/2026A&A...708A.107O}
}

@ARTICLE{SuarezMascareno2016,
       author = {{Su{\'a}rez Mascare{\~n}o}, A. and {Rebolo}, R. and {Gonz{\'a}lez Hern{\'a}ndez}, J.~I.},
        title = "{Magnetic cycles and rotation periods of late-type stars from photometric time series}",
      journal = {\aap},
         year = 2016,
        month = oct,
       volume = {595},
          eid = {A12},
        pages = {A12},
          doi = {10.1051/0004-6361/201628586},
archivePrefix = {arXiv},
       eprint = {1607.03049},
 primaryClass = {astro-ph.SR},
       adsurl = {https://ui.adsabs.harvard.edu/abs/2016A&A...595A..12S}
}

@ARTICLE{Khodachenko2007,
       author = {{Khodachenko}, Maxim L. and {Ribas}, Ignasi and {Lammer}, Helmut and {Grie{\ss}meier}, Jean-Mathias and {Leitner}, Martin and {Selsis}, Franck and {Eiroa}, Carlos and {Hanslmeier}, Arnold and {Biernat}, Helfried K. and {Farrugia}, Charles J. and {Rucker}, Helmut O.},
        title = "{Coronal Mass Ejection (CME) Activity of Low Mass M Stars as An Important Factor for The Habitability of Terrestrial Exoplanets. I. CME Impact on Expected Magnetospheres of Earth-Like Exoplanets in Close-In Habitable Zones}",
      journal = {Astrobiology},
         year = 2007,
        month = feb,
       volume = {7},
       number = {1},
        pages = {167-184},
          doi = {10.1089/ast.2006.0127},
       adsurl = {https://ui.adsabs.harvard.edu/abs/2007AsBio...7..167K}
}

@ARTICLE{Vidotto2013,
       author = {{Vidotto}, A.~A. and {Jardine}, M. and {Morin}, J. and {Donati}, J.-F. and {Lang}, P. and {Russell}, A.~J.~B.},
        title = "{Effects of M dwarf magnetic fields on potentially habitable planets}",
      journal = {\aap},
         year = 2013,
        month = sep,
       volume = {557},
          eid = {A67},
        pages = {A67},
          doi = {10.1051/0004-6361/201321504},
archivePrefix = {arXiv},
       eprint = {1306.4789},
 primaryClass = {astro-ph.EP},
       adsurl = {https://ui.adsabs.harvard.edu/abs/2013A&A...557A..67V}
}

@ARTICLE{Lanza2013,
       author = {{Lanza}, A.~F.},
        title = "{Star-planet magnetic interaction and evaporation of planetary atmospheres}",
      journal = {\aap},
         year = 2013,
        month = sep,
       volume = {557},
          eid = {A31},
        pages = {A31},
          doi = {10.1051/0004-6361/201321790},
archivePrefix = {arXiv},
       eprint = {1307.2341},
 primaryClass = {astro-ph.EP},
       adsurl = {https://ui.adsabs.harvard.edu/abs/2013A&A...557A..31L}
}

@ARTICLE{Loyd2023,
       author = {{Loyd}, R.~O. Parke and {Schneider}, P.~C. and {Jackman}, James A.~G. and {France}, Kevin and {Shkolnik}, Evgenya L. and {Arulanantham}, Nicole and {Cauley}, P. Wilson and {Llama}, Joe and {Schneider}, Adam C.},
        title = "{Flares, Rotation, Activity Cycles, and a Magnetic Star{\textendash}Planet Interaction Hypothesis for the Far-ultraviolet Emission of GJ 436}",
      journal = {\aj},
         year = 2023,
        month = apr,
       volume = {165},
       number = {4},
          eid = {146},
        pages = {146},
          doi = {10.3847/1538-3881/acbbc8},
archivePrefix = {arXiv},
       eprint = {2302.10259},
 primaryClass = {astro-ph.SR},
       adsurl = {https://ui.adsabs.harvard.edu/abs/2023AJ....165..146L}
}

@ARTICLE{Scalo2007,
       author = {{Scalo}, John and {Kaltenegger}, Lisa and {Segura}, Ant{\'\i}gona and {Fridlund}, Malcolm and {Ribas}, Ignasi and {Kulikov}, Yu. N. and {Grenfell}, John L. and {Rauer}, Heike and {Odert}, Petra and {Leitzinger}, Martin and {Selsis}, F. and {Khodachenko}, Maxim L. and {Eiroa}, Carlos and {Kasting}, Jim and {Lammer}, Helmut},
        title = "{M Stars as Targets for Terrestrial Exoplanet Searches And Biosignature Detection}",
      journal = {Astrobiology},
         year = 2007,
        month = feb,
       volume = {7},
       number = {1},
        pages = {85-166},
          doi = {10.1089/ast.2006.0125},
       adsurl = {https://ui.adsabs.harvard.edu/abs/2007AsBio...7...85S}
}

@ARTICLE{Bonfils2013,
       author = {{Bonfils}, X. and {Delfosse}, X. and {Udry}, S. and {Forveille}, T. and {Mayor}, M. and {Perrier}, C. and {Bouchy}, F. and {Gillon}, M. and {Lovis}, C. and {Pepe}, F. and {Queloz}, D. and {Santos}, N.~C. and {S{\'e}gransan}, D. and {Bertaux}, J.-L.},
        title = "{The HARPS search for southern extra-solar planets. XXXI. The M-dwarf sample}",
      journal = {\aap},
         year = 2013,
        month = jan,
       volume = {549},
          eid = {A109},
        pages = {A109},
          doi = {10.1051/0004-6361/201014704},
archivePrefix = {arXiv},
       eprint = {1111.5019},
 primaryClass = {astro-ph.EP},
       adsurl = {https://ui.adsabs.harvard.edu/abs/2013A&A...549A.109B}
}

@ARTICLE{Dressing2015,
       author = {{Dressing}, Courtney D. and {Charbonneau}, David},
        title = "{The Occurrence of Potentially Habitable Planets Orbiting M Dwarfs Estimated from the Full Kepler Dataset and an Empirical Measurement of the Detection Sensitivity}",
      journal = {\apj},
         year = 2015,
        month = jul,
       volume = {807},
       number = {1},
          eid = {45},
        pages = {45},
          doi = {10.1088/0004-637X/807/1/45},
archivePrefix = {arXiv},
       eprint = {1501.01623},
 primaryClass = {astro-ph.EP},
       adsurl = {https://ui.adsabs.harvard.edu/abs/2015ApJ...807...45D}
}

@ARTICLE{Edwards2019,
       author = {{Edwards}, Billy and {Mugnai}, Lorenzo and {Tinetti}, Giovanna and {Pascale}, Enzo and {Sarkar}, Subhajit},
        title = "{An Updated Study of Potential Targets for Ariel}",
      journal = {\aj},
         year = 2019,
        month = jun,
       volume = {157},
       number = {6},
          eid = {242},
        pages = {242},
          doi = {10.3847/1538-3881/ab1cb9},
archivePrefix = {arXiv},
       eprint = {1905.04959},
 primaryClass = {astro-ph.EP},
       adsurl = {https://ui.adsabs.harvard.edu/abs/2019AJ....157..242E}
}

@ARTICLE{Morley2017,
       author = {{Morley}, Caroline V. and {Kreidberg}, Laura and {Rustamkulov}, Zafar and {Robinson}, Tyler and {Fortney}, Jonathan J.},
        title = "{Observing the Atmospheres of Known Temperate Earth-sized Planets with JWST}",
      journal = {\apj},
         year = 2017,
        month = dec,
       volume = {850},
       number = {2},
          eid = {121},
        pages = {121},
          doi = {10.3847/1538-4357/aa927b},
archivePrefix = {arXiv},
       eprint = {1708.04239},
 primaryClass = {astro-ph.EP},
       adsurl = {https://ui.adsabs.harvard.edu/abs/2017ApJ...850..121M}
}

@ARTICLE{Tinetti2018,
       author = {{Tinetti}, Giovanna and {Drossart}, Pierre and {Eccleston}, Paul and {Hartogh}, Paul and {Heske}, Astrid and {Leconte}, J{\'e}r{\'e}my and {Micela}, Giusi and {Ollivier}, Marc and {Pilbratt}, G{\"o}ran and {Puig}, Ludovic and {Turrini}, Diego and {Vandenbussche}, Bart and {Wolkenberg}, Paulina and {Beaulieu}, Jean-Philippe and {Buchave}, Lars A. and {Ferus}, Martin and {Griffin}, Matt and {Guedel}, Manuel and {Justtanont}, Kay and {Lagage}, Pierre-Olivier and {Machado}, Pedro and {Malaguti}, Giuseppe and {Min}, Michiel and {N{\o}rgaard-Nielsen}, Hans Ulrik and {Rataj}, Mirek and {Ray}, Tom and {Ribas}, Ignasi and {Swain}, Mark and {Szabo}, Robert and {Werner}, Stephanie and {Barstow}, Joanna and {Burleigh}, Matt and {Cho}, James and {Coud{\'e} du Foresto}, Vincent and {Coustenis}, Athena and {Decin}, Leen and {Encrenaz}, Therese and {Galand}, Marina and {Gillon}, Michael and {Helled}, Ravit and {Morales}, Juan Carlos and {Garc{\'\i}a Mu{\~n}oz}, Antonio and {Moneti}, Andrea and {Pagano}, Isabella and {Pascale}, Enzo and {Piccioni}, Giuseppe and {Pinfield}, David and {Sarkar}, Subhajit and {Selsis}, Franck and {Tennyson}, Jonathan and {Triaud}, Amaury and {Venot}, Olivia and {Waldmann}, Ingo and {Waltham}, David and {Wright}, Gillian and {Amiaux}, Jerome and {Augu{\`e}res}, Jean-Louis and {Berth{\'e}}, Michel and {Bezawada}, Naidu and {Bishop}, Georgia and {Bowles}, Neil and {Coffey}, Deirdre and {Colom{\'e}}, Josep and {Crook}, Martin and {Crouzet}, Pierre-Elie and {Da Peppo}, Vania and {Sanz}, Isabel Escudero and {Focardi}, Mauro and {Frericks}, Martin and {Hunt}, Tom and {Kohley}, Ralf and {Middleton}, Kevin and {Morgante}, Gianluca and {Ottensamer}, Roland and {Pace}, Emanuele and {Pearson}, Chris and {Stamper}, Richard and {Symonds}, Kate and {Rengel}, Miriam and {Renotte}, Etienne and {Ade}, Peter and {Affer}, Laura and {Alard}, Christophe and {Allard}, Nicole and {Altieri}, Francesca and {Andr{\'e}}, Yves and {Arena}, Claudio and {Argyriou}, Ioannis and {Aylward}, Alan and {Baccani}, Cristian and {Bakos}, Gaspar and {Banaszkiewicz}, Marek and {Barlow}, Mike and {Batista}, Virginie and {Bellucci}, Giancarlo and {Benatti}, Serena and {Bernardi}, Pernelle and {B{\'e}zard}, Bruno and {Blecka}, Maria and {Bolmont}, Emeline and {Bonfond}, Bertrand and {Bonito}, Rosaria and {Bonomo}, Aldo S. and {Brucato}, John Robert and {Brun}, Allan Sacha and {Bryson}, Ian and {Bujwan}, Waldemar and {Casewell}, Sarah and {Charnay}, Bejamin and {Pestellini}, Cesare Cecchi and {Chen}, Guo and {Ciaravella}, Angela and {Claudi}, Riccardo and {Cl{\'e}dassou}, Rodolphe and {Damasso}, Mario and {Damiano}, Mario and {Danielski}, Camilla and {Deroo}, Pieter and {Di Giorgio}, Anna Maria and {Dominik}, Carsten and {Doublier}, Vanessa and {Doyle}, Simon and {Doyon}, Ren{\'e} and {Drummond}, Benjamin and {Duong}, Bastien and {Eales}, Stephen and {Edwards}, Billy and {Farina}, Maria and {Flaccomio}, Ettore and {Fletcher}, Leigh and {Forget}, Fran{\c{c}}ois and {Fossey}, Steve and {Fr{\"a}nz}, Markus and {Fujii}, Yuka and {Garc{\'\i}a-Piquer}, {\'A}lvaro and {Gear}, Walter and {Geoffray}, Herv{\'e} and {G{\'e}rard}, Jean Claude and {Gesa}, Lluis and {Gomez}, H. and {Graczyk}, Rafa{\l} and {Griffith}, Caitlin and {Grodent}, Denis and {Guarcello}, Mario Giuseppe and {Gustin}, Jacques and {Hamano}, Keiko and {Hargrave}, Peter and {Hello}, Yann and {Heng}, Kevin and {Herrero}, Enrique and {Hornstrup}, Allan and {Hubert}, Benoit and {Ida}, Shigeru and {Ikoma}, Masahiro and {Iro}, Nicolas and {Irwin}, Patrick and {Jarchow}, Christopher and {Jaubert}, Jean and {Jones}, Hugh and {Julien}, Queyrel and {Kameda}, Shingo and {Kerschbaum}, Franz and {Kervella}, Pierre and {Koskinen}, Tommi and {Krijger}, Matthijs and {Krupp}, Norbert and {Lafarga}, Marina and {Landini}, Federico and {Lellouch}, Emanuel and {Leto}, Giuseppe and {Luntzer}, A. and {Rank-L{\"u}ftinger}, Theresa and {Maggio}, Antonio and {Maldonado}, Jesus and {Maillard}, Jean-Pierre and {Mall}, Urs and {Marquette}, Jean-Baptiste and {Mathis}, Stephane and {Maxted}, Pierre and {Matsuo}, Taro and {Medvedev}, Alexander and {Miguel}, Yamila and {Minier}, Vincent and {Morello}, Giuseppe and {Mura}, Alessandro and {Narita}, Norio and {Nascimbeni}, Valerio and {Nguyen Tong}, N. and {Noce}, Vladimiro and {Oliva}, Fabrizio and {Palle}, Enric and {Palmer}, Paul and {Pancrazzi}, Maurizio and {Papageorgiou}, Andreas and {Parmentier}, Vivien and {Perger}, Manuel and {Petralia}, Antonino and {Pezzuto}, Stefano and {Pierrehumbert}, Ray and {Pillitteri}, Ignazio},
        title = "{A chemical survey of exoplanets with ARIEL}",
      journal = {Experimental Astronomy},
         year = 2018,
        month = nov,
       volume = {46},
       number = {1},
        pages = {135-209},
          doi = {10.1007/s10686-018-9598-x},
       adsurl = {https://ui.adsabs.harvard.edu/abs/2018ExA....46..135T}
}

@ARTICLE{Sabotta2021,
       author = {{Sabotta}, S. and {Schlecker}, M. and {Chaturvedi}, P. and {Guenther}, E.~W. and {Mu{\~n}oz Rodr{\'\i}guez}, I. and {Mu{\~n}oz S{\'a}nchez}, J.~C. and {Caballero}, J.~A. and {Shan}, Y. and {Reffert}, S. and {Ribas}, I. and {Reiners}, A. and {Hatzes}, A.~P. and {Amado}, P.~J. and {Klahr}, H. and {Morales}, J.~C. and {Quirrenbach}, A. and {Henning}, Th. and {Dreizler}, S. and {Pall{\'e}}, E. and {Perger}, M. and {Azzaro}, M. and {Jeffers}, S.~V. and {Kaminski}, A. and {K{\"u}rster}, M. and {Lafarga}, M. and {Montes}, D. and {Passegger}, V.~M. and {Zechmeister}, M.},
        title = "{The CARMENES search for exoplanets around M dwarfs. Planet occurrence rates from a subsample of 71 stars}",
      journal = {\aap},
         year = 2021,
        month = sep,
       volume = {653},
          eid = {A114},
        pages = {A114},
          doi = {10.1051/0004-6361/202140968},
archivePrefix = {arXiv},
       eprint = {2107.03802},
 primaryClass = {astro-ph.EP},
       adsurl = {https://ui.adsabs.harvard.edu/abs/2021A&A...653A.114S}
}

@ARTICLE{Schweitzer2019,
       author = {{Schweitzer}, A. and {Passegger}, V.~M. and {Cifuentes}, C. and {B{\'e}jar}, V.~J.~S. and {Cort{\'e}s-Contreras}, M. and {Caballero}, J.~A. and {del Burgo}, C. and {Czesla}, S. and {K{\"u}rster}, M. and {Montes}, D. and {Zapatero Osorio}, M.~R. and {Ribas}, I. and {Reiners}, A. and {Quirrenbach}, A. and {Amado}, P.~J. and {Aceituno}, J. and {Anglada-Escud{\'e}}, G. and {Bauer}, F.~F. and {Dreizler}, S. and {Jeffers}, S.~V. and {Guenther}, E.~W. and {Henning}, T. and {Kaminski}, A. and {Lafarga}, M. and {Marfil}, E. and {Morales}, J.~C. and {Schmitt}, J.~H.~M.~M. and {Seifert}, W. and {Solano}, E. and {Tabernero}, H.~M. and {Zechmeister}, M.},
        title = "{The CARMENES search for exoplanets around M dwarfs. Different roads to radii and masses of the target stars}",
      journal = {\aap},
         year = 2019,
        month = may,
       volume = {625},
          eid = {A68},
        pages = {A68},
          doi = {10.1051/0004-6361/201834965},
archivePrefix = {arXiv},
       eprint = {1904.03231},
 primaryClass = {astro-ph.SR},
       adsurl = {https://ui.adsabs.harvard.edu/abs/2019A&A...625A..68S}
}

@ARTICLE{Stock2020,
       author = {{Stock}, S. and {Kemmer}, J. and {Reffert}, S. and {Trifonov}, T. and {Kaminski}, A. and {Dreizler}, S. and {Quirrenbach}, A. and {Caballero}, J.~A. and {Reiners}, A. and {Jeffers}, S.~V. and {Anglada-Escud{\'e}}, G. and {Ribas}, I. and {Amado}, P.~J. and {Barrado}, D. and {Barnes}, J.~R. and {Bauer}, F.~F. and {Berdi{\~n}as}, Z.~M. and {B{\'e}jar}, V.~J.~S. and {Coleman}, G.~A.~L. and {Cort{\'e}s-Contreras}, M. and {D{\'\i}ez-Alonso}, E. and {Dom{\'\i}nguez-Fern{\'a}ndez}, A.~J. and {Espinoza}, N. and {Haswell}, C.~A. and {Hatzes}, A. and {Henning}, T. and {Jenkins}, J.~S. and {Jones}, H.~R.~A. and {Kossakowski}, D. and {K{\"u}rster}, M. and {Lafarga}, M. and {Lee}, M.~H. and {L{\'o}pez Gonz{\'a}lez}, M.~J. and {Montes}, D. and {Morales}, J.~C. and {Morales}, N. and {Pall{\'e}}, E. and {Pedraz}, S. and {Rodr{\'\i}guez}, E. and {Rodr{\'\i}guez-L{\'o}pez}, C. and {Zechmeister}, M.},
        title = "{The CARMENES search for exoplanets around M dwarfs. Characterization of the nearby ultra-compact multiplanetary system YZ Ceti}",
      journal = {\aap},
         year = 2020,
        month = apr,
       volume = {636},
          eid = {A119},
        pages = {A119},
          doi = {10.1051/0004-6361/201936732},
archivePrefix = {arXiv},
       eprint = {2002.01772},
 primaryClass = {astro-ph.EP},
       adsurl = {https://ui.adsabs.harvard.edu/abs/2020A&A...636A.119S}
}

@ARTICLE{Donati2023,
       author = {{Donati}, J.-F. and {Lehmann}, L.~T. and {Cristofari}, P.~I. and {Fouqu{\'e}}, P. and {Moutou}, C. and {Charpentier}, P. and {Ould-Elhkim}, M. and {Carmona}, A. and {Delfosse}, X. and {Artigau}, E. and {Alencar}, S.~H.~P. and {Cadieux}, C. and {Arnold}, L. and {Petit}, P. and {Morin}, J. and {Forveille}, T. and {Cloutier}, R. and {Doyon}, R. and {H{\'e}brard}, G. and {SLS Collaboration}},
        title = "{Magnetic fields and rotation periods of M dwarfs from SPIRou spectra}",
      journal = {\mnras},
         year = 2023,
        month = oct,
       volume = {525},
       number = {2},
        pages = {2015-2039},
          doi = {10.1093/mnras/stad2301},
archivePrefix = {arXiv},
       eprint = {2307.14190},
 primaryClass = {astro-ph.SR},
       adsurl = {https://ui.adsabs.harvard.edu/abs/2023MNRAS.525.2015D}
}

@ARTICLE{Cristofari2022,
       author = {{Cristofari}, P.~I. and {Donati}, J.-F. and {Masseron}, T. and {Fouqu{\'e}}, P. and {Moutou}, C. and {Carmona}, A. and {Artigau}, E. and {Martioli}, E. and {H{\'e}brard}, G. and {Gaidos}, E. and {Delfosse}, X. and {SLS consortium}},
        title = "{Estimating the atmospheric properties of 44 M dwarfs from SPIRou spectra}",
      journal = {\mnras},
         year = 2022,
        month = nov,
       volume = {516},
       number = {3},
        pages = {3802-3820},
          doi = {10.1093/mnras/stac2364},
archivePrefix = {arXiv},
       eprint = {2208.10340},
 primaryClass = {astro-ph.SR},
       adsurl = {https://ui.adsabs.harvard.edu/abs/2022MNRAS.516.3802C}
}

@ARTICLE{Chebly2023,
       author = {{Chebly}, Judy J. and {Alvarado-G{\'o}mez}, Juli{\'a}n D. and {Poppenh{\"a}ger}, Katja and {Garraffo}, Cecilia},
        title = "{Numerical quantification of the wind properties of cool main sequence stars}",
      journal = {\mnras},
         year = 2023,
        month = oct,
       volume = {524},
       number = {4},
        pages = {5060-5079},
          doi = {10.1093/mnras/stad2100},
archivePrefix = {arXiv},
       eprint = {2307.04615},
 primaryClass = {astro-ph.SR},
       adsurl = {https://ui.adsabs.harvard.edu/abs/2023MNRAS.524.5060C}
}

@ARTICLE{Vidotto2017,
       author = {{Vidotto}, A.~A. and {Bourrier}, V.},
        title = "{Exoplanets as probes of the winds of host stars: the case of the M dwarf GJ 436}",
      journal = {\mnras},
         year = 2017,
        month = oct,
       volume = {470},
       number = {4},
        pages = {4026-4033},
          doi = {10.1093/mnras/stx1543},
archivePrefix = {arXiv},
       eprint = {1706.05894},
 primaryClass = {astro-ph.SR},
       adsurl = {https://ui.adsabs.harvard.edu/abs/2017MNRAS.470.4026V}
}

@ARTICLE{Jardine2019,
       author = {{Jardine}, Moira and {Collier Cameron}, Andrew},
        title = "{Slingshot prominences: nature's wind gauges}",
      journal = {\mnras},
         year = 2019,
        month = jan,
       volume = {482},
       number = {3},
        pages = {2853-2860},
          doi = {10.1093/mnras/sty2872},
archivePrefix = {arXiv},
       eprint = {1810.09319},
 primaryClass = {astro-ph.SR},
       adsurl = {https://ui.adsabs.harvard.edu/abs/2019MNRAS.482.2853J}
}

@ARTICLE{See2017,
       author = {{See}, V. and {Jardine}, M. and {Vidotto}, A.~A. and {Donati}, J.-F. and {Boro Saikia}, S. and {Fares}, R. and {Folsom}, C.~P. and {H{\'e}brard}, {\'E}. M. and {Jeffers}, S.~V. and {Marsden}, S.~C. and {Morin}, J. and {Petit}, P. and {Waite}, I.~A. and {BCool Collaboration}},
        title = "{Studying stellar spin-down with Zeeman-Doppler magnetograms}",
      journal = {\mnras},
         year = 2017,
        month = apr,
       volume = {466},
       number = {2},
        pages = {1542-1554},
          doi = {10.1093/mnras/stw3094},
archivePrefix = {arXiv},
       eprint = {1705.01835},
 primaryClass = {astro-ph.SR},
       adsurl = {https://ui.adsabs.harvard.edu/abs/2017MNRAS.466.1542S}
}

@ARTICLE{Wood2021,
       author = {{Wood}, Brian E. and {M{\"u}ller}, Hans-Reinhard and {Redfield}, Seth and {Konow}, Fallon and {Vannier}, Hunter and {Linsky}, Jeffrey L. and {Youngblood}, Allison and {Vidotto}, Aline A. and {Jardine}, Moira and {Alvarado-G{\'o}mez}, Juli{\'a}n D. and {Drake}, Jeremy J.},
        title = "{New Observational Constraints on the Winds of M dwarf Stars}",
      journal = {\apj},
         year = 2021,
        month = jul,
       volume = {915},
       number = {1},
          eid = {37},
        pages = {37},
          doi = {10.3847/1538-4357/abfda5},
archivePrefix = {arXiv},
       eprint = {2105.00019},
 primaryClass = {astro-ph.SR},
       adsurl = {https://ui.adsabs.harvard.edu/abs/2021ApJ...915...37W}
}

@ARTICLE{Evensberget2023,
       author = {{Evensberget}, D. and {Marsden}, S.~C. and {Carter}, B.~D. and {Salmeron}, R. and {Vidotto}, A.~A. and {Folsom}, C.~P. and {Kavanagh}, R.~D. and {Pineda}, J.~S. and {Driessen}, F.~A. and {Strickert}, K.~M.},
        title = "{The winds of young Solar-type stars in the Pleiades, AB Doradus, Columba, and {\ensuremath{\beta}} Pictoris}",
      journal = {\mnras},
         year = 2023,
        month = sep,
       volume = {524},
       number = {2},
        pages = {2042-2063},
          doi = {10.1093/mnras/stad1650},
archivePrefix = {arXiv},
       eprint = {2305.17427},
 primaryClass = {astro-ph.SR},
       adsurl = {https://ui.adsabs.harvard.edu/abs/2023MNRAS.524.2042E}
}

@ARTICLE{Evensberget2022,
       author = {{Evensberget}, D. and {Carter}, B.~D. and {Marsden}, S.~C. and {Brookshaw}, L. and {Folsom}, C.~P. and {Salmeron}, R.},
        title = "{The winds of young Solar-type stars in Coma Berenices and Hercules-Lyra}",
      journal = {\mnras},
         year = 2022,
        month = mar,
       volume = {510},
       number = {4},
        pages = {5226-5245},
          doi = {10.1093/mnras/stab3557},
archivePrefix = {arXiv},
       eprint = {2112.01445},
 primaryClass = {astro-ph.SR},
       adsurl = {https://ui.adsabs.harvard.edu/abs/2022MNRAS.510.5226E}
}

@ARTICLE{Vidotto2021,
       author = {{Vidotto}, Aline A.},
        title = "{The evolution of the solar wind}",
      journal = {Living Reviews in Solar Physics},
         year = 2021,
        month = dec,
       volume = {18},
       number = {1},
          eid = {3},
        pages = {3},
          doi = {10.1007/s41116-021-00029-w},
archivePrefix = {arXiv},
       eprint = {2103.15748},
 primaryClass = {astro-ph.SR},
       adsurl = {https://ui.adsabs.harvard.edu/abs/2021LRSP...18....3V}
}

@ARTICLE{Hebrard2016,
       author = {{H{\'e}brard}, {\'E}. M. and {Donati}, J.-F. and {Delfosse}, X. and {Morin}, J. and {Moutou}, C. and {Boisse}, I.},
        title = "{Modelling the RV jitter of early-M dwarfs using tomographic imaging}",
      journal = {\mnras},
         year = 2016,
        month = sep,
       volume = {461},
       number = {2},
        pages = {1465-1497},
          doi = {10.1093/mnras/stw1346},
archivePrefix = {arXiv},
       eprint = {1606.01775},
 primaryClass = {astro-ph.SR},
       adsurl = {https://ui.adsabs.harvard.edu/abs/2016MNRAS.461.1465H}
}

@ARTICLE{Smith2026,
       author = {{Smith}, K.~G. and {Evensberget}, D. and {Bellotti}, S. and {Morin}, J. and {Vidotto}, A.~A. and {Pope}, B.~J.~S.},
        title = "{The attempted polarity reversal and evolving magnetic environment of AD Leo}",
      journal = {\aap},
         year = 2026,
        month = mar,
       volume = {707},
          eid = {A397},
        pages = {A397},
          doi = {10.1051/0004-6361/202555330},
archivePrefix = {arXiv},
       eprint = {2602.18085},
 primaryClass = {astro-ph.SR},
       adsurl = {https://ui.adsabs.harvard.edu/abs/2026A&A...707A.397S}
}

@ARTICLE{Vidotto2025,
       author = {{Vidotto}, A.~A.},
        title = "{Star{\textendash}Planet Interactions: A Computational View}",
      journal = {\araa},
         year = 2025,
        month = aug,
       volume = {63},
       number = {1},
        pages = {299-337},
          doi = {10.1146/annurev-astro-021225-030604},
archivePrefix = {arXiv},
       eprint = {2506.00470},
 primaryClass = {astro-ph.EP},
       adsurl = {https://ui.adsabs.harvard.edu/abs/2025ARA&A..63..299V}
}

@ARTICLE{Vidotto2023,
       author = {{Vidotto}, A.~A. and {Bourrier}, V. and {Fares}, R. and {Bellotti}, S. and {Donati}, J.~F. and {Petit}, P. and {Hussain}, G.~A.~J. and {Morin}, J.},
        title = "{The space weather around the exoplanet GJ 436b. II. Stellar wind-exoplanet interactions}",
      journal = {\aap},
         year = 2023,
        month = oct,
       volume = {678},
          eid = {A152},
        pages = {A152},
          doi = {10.1051/0004-6361/202347237},
archivePrefix = {arXiv},
       eprint = {2309.00324},
 primaryClass = {astro-ph.SR},
       adsurl = {https://ui.adsabs.harvard.edu/abs/2023A&A...678A.152V}
}

@ARTICLE{Dorman1989,
       author = {{Dorman}, Ben and {Nelson}, Lorne A. and {Chau}, W.~Y.},
        title = "{Theoretical Models of Low-Mass Stars and Brown Dwarfs. I. The Lower Main Sequence}",
      journal = {\apj},
         year = 1989,
        month = jul,
       volume = {342},
        pages = {1003},
          doi = {10.1086/167658},
       adsurl = {https://ui.adsabs.harvard.edu/abs/1989ApJ...342.1003D}
}

@ARTICLE{Rabus2019,
       author = {{Rabus}, Markus and {Lachaume}, R{\'e}gis and {Jord{\'a}n}, Andr{\'e}s and {Brahm}, Rafael and {Boyajian}, Tabetha and {von Braun}, Kaspar and {Espinoza}, N{\'e}stor and {Berger}, Jean-Philippe and {Le Bouquin}, Jean-Baptiste and {Absil}, Olivier},
        title = "{A discontinuity in the T$_{eff}$-radius relation of M-dwarfs}",
      journal = {\mnras},
         year = 2019,
        month = apr,
       volume = {484},
       number = {2},
        pages = {2674-2683},
          doi = {10.1093/mnras/sty3430},
archivePrefix = {arXiv},
       eprint = {1901.08077},
 primaryClass = {astro-ph.SR},
       adsurl = {https://ui.adsabs.harvard.edu/abs/2019MNRAS.484.2674R}
}

@ARTICLE{Donati2020,
       author = {{Donati}, J.-F. and {Kouach}, D. and {Moutou}, C. and {Doyon}, R. and {Delfosse}, X. and {Artigau}, E. and {Baratchart}, S. and {Lacombe}, M. and {Barrick}, G. and {H{\'e}brard}, G. and {Bouchy}, F. and {Saddlemyer}, L. and {Par{\`e}s}, L. and {Rabou}, P. and {Micheau}, Y. and {Dolon}, F. and {Reshetov}, V. and {Challita}, Z. and {Carmona}, A. and {Striebig}, N. and {Thibault}, S. and {Martioli}, E. and {Cook}, N. and {Fouqu{\'e}}, P. and {Vermeulen}, T. and {Wang}, S.~Y. and {Arnold}, L. and {Pepe}, F. and {Boisse}, I. and {Figueira}, P. and {Bouvier}, J. and {Ray}, T.~P. and {Feugeade}, C. and {Morin}, J. and {Alencar}, S. and {Hobson}, M. and {Castilho}, B. and {Udry}, S. and {Santos}, N.~C. and {Hernandez}, O. and {Benedict}, T. and {Vall{\'e}e}, P. and {Gallou}, G. and {Dupieux}, M. and {Larrieu}, M. and {Perruchot}, S. and {Sottile}, R. and {Moreau}, F. and {Usher}, C. and {Baril}, M. and {Wildi}, F. and {Chazelas}, B. and {Malo}, L. and {Bonfils}, X. and {Loop}, D. and {Kerley}, D. and {Wevers}, I. and {Dunn}, J. and {Pazder}, J. and {Macdonald}, S. and {Dubois}, B. and {Carri{\'e}}, E. and {Valentin}, H. and {Henault}, F. and {Yan}, C.~H. and {Steinmetz}, T.},
        title = "{SPIRou: NIR velocimetry and spectropolarimetry at the CFHT}",
      journal = {\mnras},
         year = 2020,
        month = nov,
       volume = {498},
       number = {4},
        pages = {5684-5703},
          doi = {10.1093/mnras/staa2569},
archivePrefix = {arXiv},
       eprint = {2008.08949},
 primaryClass = {astro-ph.IM},
       adsurl = {https://ui.adsabs.harvard.edu/abs/2020MNRAS.498.5684D}
}

@ARTICLE{Wright2018,
       author = {{Wright}, Nicholas J. and {Newton}, Elisabeth R. and {Williams}, Peter K.~G. and {Drake}, Jeremy J. and {Yadav}, Rakesh K.},
        title = "{The stellar rotation-activity relationship in fully convective M dwarfs}",
      journal = {\mnras},
         year = 2018,
        month = sep,
       volume = {479},
       number = {2},
        pages = {2351-2360},
          doi = {10.1093/mnras/sty1670},
archivePrefix = {arXiv},
       eprint = {1807.03304},
 primaryClass = {astro-ph.SR},
       adsurl = {https://ui.adsabs.harvard.edu/abs/2018MNRAS.479.2351W}
}

@ARTICLE{Chabrier1997,
       author = {{Chabrier}, Gilles and {Baraffe}, Isabelle},
        title = "{Structure and evolution of low-mass stars}",
      journal = {\aap},
         year = 1997,
        month = nov,
       volume = {327},
        pages = {1039-1053},
          doi = {10.48550/arXiv.astro-ph/9704118},
archivePrefix = {arXiv},
       eprint = {astro-ph/9704118},
 primaryClass = {astro-ph},
       adsurl = {https://ui.adsabs.harvard.edu/abs/1997A&A...327.1039C}
}

@ARTICLE{Callingham2025,
       author = {{Callingham}, J.~R. and {Tasse}, C. and {Keers}, R. and {Kavanagh}, R.~D. and {Vedantham}, H.~K. and {Zarka}, P. and {Bellotti}, S. and {Cristofari}, P.~I. and {Bloot}, S. and {Konijn}, D.~C. and {Hardcastle}, M.~J. and {Lamy}, L. and {Pass}, E.~K. and {Pope}, B.~J.~S. and {Reid}, H. and {R{\"o}ttgering}, H.~J.~A. and {Shimwell}, T.~W. and {Zucca}, P.},
        title = "{Radio burst from a stellar coronal mass ejection}",
      journal = {\nat},
         year = 2025,
        month = nov,
       volume = {647},
       number = {8090},
        pages = {603-607},
          doi = {10.1038/s41586-025-09715-3},
archivePrefix = {arXiv},
       eprint = {2511.09289},
 primaryClass = {astro-ph.SR},
       adsurl = {https://ui.adsabs.harvard.edu/abs/2025Natur.647..603C}
}

@ARTICLE{Charbonneau2010,
       author = {{Charbonneau}, Paul},
        title = "{Dynamo Models of the Solar Cycle}",
      journal = {Living Reviews in Solar Physics},
         year = 2010,
        month = dec,
       volume = {7},
       number = {1},
          eid = {3},
        pages = {3},
          doi = {10.12942/lrsp-2010-3},
       adsurl = {https://ui.adsabs.harvard.edu/abs/2010LRSP....7....3C}
}

@ARTICLE{WeberDavis1967,
       author = {{Weber}, Edmund J. and {Davis}, Jr., Leverett},
        title = "{The Angular Momentum of the Solar Wind}",
      journal = {\apj},
         year = 1967,
        month = apr,
       volume = {148},
        pages = {217-227},
          doi = {10.1086/149138},
       adsurl = {https://ui.adsabs.harvard.edu/abs/1967ApJ...148..217W}
}

@ARTICLE{Vidotto2014b,
       author = {{Vidotto}, A.~A. and {Jardine}, M. and {Morin}, J. and {Donati}, J.~F. and {Opher}, M. and {Gombosi}, T.~I.},
        title = "{M-dwarf stellar winds: the effects of realistic magnetic geometry on rotational evolution and planets}",
      journal = {\mnras},
         year = 2014,
        month = feb,
       volume = {438},
       number = {2},
        pages = {1162-1175},
          doi = {10.1093/mnras/stt2265},
archivePrefix = {arXiv},
       eprint = {1311.5063},
 primaryClass = {astro-ph.SR},
       adsurl = {https://ui.adsabs.harvard.edu/abs/2014MNRAS.438.1162V}
}

@ARTICLE{SciPy,
  author  = {Virtanen, Pauli and Gommers, Ralf and Oliphant, Travis E. and
            Haberland, Matt and Reddy, Tyler and Cournapeau, David and
            Burovski, Evgeni and Peterson, Pearu and Weckesser, Warren and
            Bright, Jonathan and {van der Walt}, St{\'e}fan J. and
            Brett, Matthew and Wilson, Joshua and Millman, K. Jarrod and
            Mayorov, Nikolay and Nelson, Andrew R. J. and Jones, Eric and
            Kern, Robert and Larson, Eric and Carey, C J and
            Polat, {\.I}lhan and Feng, Yu and Moore, Eric W. and
            {VanderPlas}, Jake and Laxalde, Denis and Perktold, Josef and
            Cimrman, Robert and Henriksen, Ian and Quintero, E. A. and
            Harris, Charles R. and Archibald, Anne M. and
            Ribeiro, Ant{\^o}nio H. and Pedregosa, Fabian and
            {van Mulbregt}, Paul and {SciPy 1.0 Contributors}},
  title   = {{{SciPy} 1.0: Fundamental Algorithms for Scientific
            Computing in Python}},
  journal = {Nature Methods},
  year    = {2020},
  volume  = {17},
  pages   = {261--272},
  adsurl  = {https://rdcu.be/b08Wh},
  doi     = {10.1038/s41592-019-0686-2},
}

@Article{NumPy,
 title         = {Array programming with {NumPy}},
 author        = {Charles R. Harris and K. Jarrod Millman and St{\'{e}}fan J.
                 van der Walt and Ralf Gommers and Pauli Virtanen and David
                 Cournapeau and Eric Wieser and Julian Taylor and Sebastian
                 Berg and Nathaniel J. Smith and Robert Kern and Matti Picus
                 and Stephan Hoyer and Marten H. van Kerkwijk and Matthew
                 Brett and Allan Haldane and Jaime Fern{\'{a}}ndez del
                 R{\'{i}}o and Mark Wiebe and Pearu Peterson and Pierre
                 G{\'{e}}rard-Marchant and Kevin Sheppard and Tyler Reddy and
                 Warren Weckesser and Hameer Abbasi and Christoph Gohlke and
                 Travis E. Oliphant},
 year          = {2020},
 month         = sep,
 journal       = {Nature},
 volume        = {585},
 number        = {7825},
 pages         = {357--362},
 doi           = {10.1038/s41586-020-2649-2},
 publisher     = {Springer Science and Business Media {LLC}},
 url           = {https://doi.org/10.1038/s41586-020-2649-2}
}

@ARTICLE{BoroSaikia2020,
       author = {{Boro Saikia}, S. and {Jin}, M. and {Johnstone}, C.~P. and {L{\"u}ftinger}, T. and {G{\"u}del}, M. and {Airapetian}, V.~S. and {Kislyakova}, K.~G. and {Folsom}, C.~P.},
        title = "{The solar wind from a stellar perspective. How do low-resolution data impact the determination of wind properties?}",
      journal = {\aap},
         year = 2020,
        month = mar,
       volume = {635},
          eid = {A178},
        pages = {A178},
          doi = {10.1051/0004-6361/201937107},
archivePrefix = {arXiv},
       eprint = {2002.10460},
 primaryClass = {astro-ph.SR},
       adsurl = {https://ui.adsabs.harvard.edu/abs/2020A&A...635A.178B}
}

@ARTICLE{Evensberget2021,
       author = {{Evensberget}, D. and {Carter}, B.~D. and {Marsden}, S.~C. and {Brookshaw}, L. and {Folsom}, C.~P.},
        title = "{The winds of young Solar-type stars in the Hyades}",
      journal = {\mnras},
         year = 2021,
        month = sep,
       volume = {506},
       number = {2},
        pages = {2309-2335},
          doi = {10.1093/mnras/stab1696},
archivePrefix = {arXiv},
       eprint = {2106.04937},
 primaryClass = {astro-ph.SR},
       adsurl = {https://ui.adsabs.harvard.edu/abs/2021MNRAS.506.2309E}
}

@ARTICLE{Sokolov2013,
       author = {{Sokolov}, Igor V. and {van der Holst}, Bart and {Oran}, Rona and {Downs}, Cooper and {Roussev}, Ilia I. and {Jin}, Meng and {Manchester}, IV, Ward B. and {Evans}, Rebekah M. and {Gombosi}, Tamas I.},
        title = "{Magnetohydrodynamic Waves and Coronal Heating: Unifying Empirical and MHD Turbulence Models}",
      journal = {\apj},
         year = 2013,
        month = feb,
       volume = {764},
       number = {1},
          eid = {23},
        pages = {23},
          doi = {10.1088/0004-637X/764/1/23},
archivePrefix = {arXiv},
       eprint = {1208.3141},
 primaryClass = {astro-ph.SR},
       adsurl = {https://ui.adsabs.harvard.edu/abs/2013ApJ...764...23S}
}

@ARTICLE{vanderHolst2014,
       author = {{van der Holst}, B. and {Sokolov}, I.~V. and {Meng}, X. and {Jin}, M. and {Manchester}, IV, W.~B. and {T{\'o}th}, G. and {Gombosi}, T.~I.},
        title = "{Alfv{\'e}n Wave Solar Model (AWSoM): Coronal Heating}",
      journal = {\apj},
         year = 2014,
        month = feb,
       volume = {782},
       number = {2},
          eid = {81},
        pages = {81},
          doi = {10.1088/0004-637X/782/2/81},
archivePrefix = {arXiv},
       eprint = {1311.4093},
 primaryClass = {astro-ph.SR},
       adsurl = {https://ui.adsabs.harvard.edu/abs/2014ApJ...782...81V}
}

@ARTICLE{Toth2005,
       author = {{T{\'o}th}, G{\'a}Bor and {Sokolov}, Igor V. and {Gombosi}, Tamas I. and {Chesney}, David R. and {Clauer}, C. Robert and {de Zeeuw}, Darren L. and {Hansen}, Kenneth C. and {Kane}, Kevin J. and {Manchester}, Ward B. and {Oehmke}, Robert C. and {Powell}, Kenneth G. and {Ridley}, Aaron J. and {Roussev}, Ilia I. and {Stout}, Quentin F. and {Volberg}, Ovsei and {Wolf}, Richard A. and {Sazykin}, Stanislav and {Chan}, Anthony and {Yu}, Bin and {K{\'o}ta}, J{\'o}zsef},
        title = "{Space Weather Modeling Framework: A new tool for the space science community}",
      journal = {Journal of Geophysical Research (Space Physics)},
         year = 2005,
        month = dec,
       volume = {110},
       number = {A12},
          eid = {A12226},
        pages = {A12226},
          doi = {10.1029/2005JA011126},
       adsurl = {https://ui.adsabs.harvard.edu/abs/2005JGRA..11012226T}
}

@ARTICLE{Toth2012,
       author = {{T{\'o}th}, G{\'a}bor and {van der Holst}, Bart and {Sokolov}, Igor V. and {De Zeeuw}, Darren L. and {Gombosi}, Tamas I. and {Fang}, Fang and {Manchester}, Ward B. and {Meng}, Xing and {Najib}, Dalal and {Powell}, Kenneth G. and {Stout}, Quentin F. and {Glocer}, Alex and {Ma}, Ying-Juan and {Opher}, Merav},
        title = "{Adaptive numerical algorithms in space weather modeling}",
      journal = {Journal of Computational Physics},
         year = 2012,
        month = feb,
       volume = {231},
       number = {3},
        pages = {870-903},
          doi = {10.1016/j.jcp.2011.02.006},
       adsurl = {https://ui.adsabs.harvard.edu/abs/2012JCoPh.231..870T}
}

@ARTICLE{Blanco-Pozo2023,
       author = {{Blanco-Pozo}, J. and {Perger}, M. and {Damasso}, M. and {Anglada Escud{\'e}}, G. and {Ribas}, I. and {Baroch}, D. and {Caballero}, J.~A. and {Cifuentes}, C. and {Jeffers}, S.~V. and {Lafarga}, M. and {Kaminski}, A. and {Kaur}, S. and {Nagel}, E. and {Perdelwitz}, V. and {P{\'e}rez-Torres}, M. and {Sozzetti}, A. and {Vigan{\`o}}, D. and {Amado}, P.~J. and {Andreuzzi}, G. and {B{\'e}jar}, V.~J.~S. and {Brown}, E.~L. and {Del Sordo}, F. and {Dreizler}, S. and {Galad{\'\i}-Enr{\'\i}quez}, D. and {Hatzes}, A.~P. and {K{\"u}rster}, M. and {Lanza}, A.~F. and {Melis}, A. and {Molinari}, E. and {Montes}, D. and {Murgia}, M. and {Pall{\'e}}, E. and {Pe{\~n}a-Mo{\~n}ino}, L. and {Perrodin}, D. and {Pilia}, M. and {Poretti}, E. and {Quirrenbach}, A. and {Reiners}, A. and {Schweitzer}, A. and {Zapatero Osorio}, M.~R. and {Zechmeister}, M.},
        title = "{The CARMENES search for exoplanets around M dwarfs. A long-period planet around GJ 1151 measured with CARMENES and HARPS-N data}",
      journal = {\aap},
         year = 2023,
        month = mar,
       volume = {671},
          eid = {A50},
        pages = {A50},
          doi = {10.1051/0004-6361/202245053},
archivePrefix = {arXiv},
       eprint = {2301.04442},
 primaryClass = {astro-ph.EP},
       adsurl = {https://ui.adsabs.harvard.edu/abs/2023A&A...671A..50B}
}

@ARTICLE{Astudillo-Defru2017,
       author = {{Astudillo-Defru}, N. and {D{\'\i}az}, R.~F. and {Bonfils}, X. and {Almenara}, J.~M. and {Delisle}, J.-B. and {Bouchy}, F. and {Delfosse}, X. and {Forveille}, T. and {Lovis}, C. and {Mayor}, M. and {Murgas}, F. and {Pepe}, F. and {Santos}, N.~C. and {S{\'e}gransan}, D. and {Udry}, S. and {W{\"u}nsche}, A.},
        title = "{The HARPS search for southern extra-solar planets. XLII. A system of Earth-mass planets around the nearby M dwarf YZ Ceti}",
      journal = {\aap},
         year = 2017,
        month = sep,
       volume = {605},
          eid = {L11},
        pages = {L11},
          doi = {10.1051/0004-6361/201731581},
archivePrefix = {arXiv},
       eprint = {1708.03336},
 primaryClass = {astro-ph.EP},
       adsurl = {https://ui.adsabs.harvard.edu/abs/2017A&A...605L..11A}
}

@ARTICLE{Carmona2025,
       author = {{Carmona}, A. and {Delfosse}, X. and {Ould-Elhkim}, M. and {Cort{\'e}s-Zuleta}, P. and {Hara}, N.~C. and {Artigau}, E. and {Moutou}, C. and {Petit}, A.~C. and {Mignon}, L. and {Donati}, J.~F. and {Cook}, N.~J. and {Gagn{\'e}}, J. and {Forveille}, T. and {Diaz}, R.~F. and {Martioli}, E. and {Arnold}, L. and {Cadieux}, C. and {Boisse}, I. and {Morin}, J. and {Petit}, P. and {Fouqu{\'e}}, P. and {Bonfils}, X. and {H{\'e}brard}, G. and {Acu{\~n}a}, L. and {do Nascimento}, Jr., J.-D. and {Cloutier}, R. and {Astudillo-Defru}, N. and {Bouchy}, F. and {Bourrier}, V. and {Dalal}, S. and {Deleuil}, M. and {Doyon}, R. and {Dumusque}, X. and {Grouffal}, S. and {Heidari}, N. and {Hoyer}, S. and {Larue}, P. and {Kiefer}, F. and {Santerne}, A. and {S{\'e}gransan}, D. and {Serrano Bell}, J. and {Stalport}, M. and {Sulis}, S. and {Udry}, S. and {Vivien}, H.~G.},
        title = "{Characterizing planetary systems with SPIRou: Detection of a sub-Neptune in a 6-day period orbit around the M dwarf Gl 410}",
      journal = {\aap},
         year = 2025,
        month = aug,
       volume = {700},
          eid = {A222},
        pages = {A222},
          doi = {10.1051/0004-6361/202452513},
archivePrefix = {arXiv},
       eprint = {2504.03572},
 primaryClass = {astro-ph.EP},
       adsurl = {https://ui.adsabs.harvard.edu/abs/2025A&A...700A.222C}
}

@ARTICLE{Biswas2025,
       author = {{Biswas}, Ayan and {Folsom}, Colin P. and {Barron}, James A. and {Wade}, Gregg A. and {Bellotti}, Stefano and {Trigilio}, Corrado},
        title = "{Discovery of a Rapidly Evolving Global Magnetic Field in the M Dwarf YZ Cet and Constraints on the Magnetic Field of Its Planet YZ Cet b}",
      journal = {\apj},
         year = 2026,
        month = feb,
       volume = {997},
       number = {2},
          eid = {268},
        pages = {268},
          doi = {10.3847/1538-4357/ae2fb4},
archivePrefix = {arXiv},
       eprint = {2512.16298},
 primaryClass = {astro-ph.SR},
       adsurl = {https://ui.adsabs.harvard.edu/abs/2026ApJ...997..268B}
}

@ARTICLE{Lehmann2024,
       author = {{Lehmann}, L.~T. and {Donati}, J.-F. and {Fouqu{\'e}}, P. and {Moutou}, C. and {Bellotti}, S. and {Delfosse}, X. and {Petit}, P. and {Carmona}, A. and {Morin}, J. and {Vidotto}, A.~A. and {the SLS consortium}},
        title = "{SPIRou reveals unusually strong magnetic fields of slowly rotating M dwarfs}",
      journal = {\mnras},
         year = 2024,
        month = jan,
       volume = {527},
       number = {2},
        pages = {4330-4352},
          doi = {10.1093/mnras/stad3472},
archivePrefix = {arXiv},
       eprint = {2311.05039},
 primaryClass = {astro-ph.SR},
       adsurl = {https://ui.adsabs.harvard.edu/abs/2024MNRAS.527.4330L}
}

@ARTICLE{Bellotti2023,
       author = {{Bellotti}, S. and {Morin}, J. and {Lehmann}, L.~T. and {Folsom}, C.~P. and {Hussain}, G.~A.~J. and {Petit}, P. and {Donati}, J.-F. and {Lavail}, A. and {Carmona}, A. and {Martioli}, E. and {Romano Zaire}, B. and {Alecian}, E. and {Moutou}, C. and {Fouqu{\'e}}, P. and {Alencar}, S. and {Artigau}, E. and {Boisse}, I. and {Bouchy}, F. and {Cadieux}, C. and {Cloutier}, R. and {Cook}, N.~J. and {Delfosse}, X. and {Doyon}, R. and {H{\'e}brard}, G. and {Kochukhov}, O. and {Wade}, G.~A.},
        title = "{Monitoring the large-scale magnetic field of AD Leo with SPIRou, ESPaDOnS, and Narval. Towards a magnetic polarity reversal?}",
      journal = {\aap},
         year = 2023,
        month = aug,
       volume = {676},
          eid = {A56},
        pages = {A56},
          doi = {10.1051/0004-6361/202346845},
archivePrefix = {arXiv},
       eprint = {2307.01016},
 primaryClass = {astro-ph.SR},
       adsurl = {https://ui.adsabs.harvard.edu/abs/2023A&A...676A..56B}
}

@ARTICLE{Donati1997_LSD,
       author = {{Donati}, J.-F. and {Semel}, M. and {Carter}, B.~D. and {Rees}, D.~E. and {Collier Cameron}, A.},
        title = "{Spectropolarimetric observations of active stars}",
      journal = {\mnras},
         year = 1997,
        month = nov,
       volume = {291},
       number = {4},
        pages = {658-682},
          doi = {10.1093/mnras/291.4.658},
       adsurl = {https://ui.adsabs.harvard.edu/abs/1997MNRAS.291..658D}
}

@ARTICLE{DonatiBrown1997,
       author = {{Donati}, J.-F. and {Brown}, S.~F.},
        title = "{Zeeman-Doppler imaging of active stars. V. Sensitivity of maximum entropy magnetic maps to field orientation.}",
      journal = {\aap},
         year = 1997,
        month = oct,
       volume = {326},
        pages = {1135-1142},
       adsurl = {https://ui.adsabs.harvard.edu/abs/1997A&A...326.1135D}
}

@ARTICLE{Semel1989,
       author = {{Semel}, M.},
        title = "{Zeeman-Doppler imaging of active stars. I - Basic principles.}",
      journal = {\aap},
         year = 1989,
        month = nov,
       volume = {225},
        pages = {456-466},
       adsurl = {https://ui.adsabs.harvard.edu/abs/1989A&A...225..456S}
}

@ARTICLE{Bellotti2025a,
       author = {{Bellotti}, S. and {Cristofari}, P.~I. and {Callingham}, J.~R. and {Morin}, J. and {Petit}, P. and {Vidotto}, A.~A. and {Jardine}, M. and {Arnold}, L. and {Kavanagh}, R.~D. and {LLama}, J. and {Vedantham}, H.},
        title = "{A magnetic field study of two fast-rotating, radio-bright M dwarfs: StKM 1-1262 and V374 Peg}",
      journal = {\aap},
         year = 2025,
        month = dec,
       volume = {704},
          eid = {A298},
        pages = {A298},
          doi = {10.1051/0004-6361/202556787},
archivePrefix = {arXiv},
       eprint = {2511.09312},
 primaryClass = {astro-ph.SR},
       adsurl = {https://ui.adsabs.harvard.edu/abs/2025A&A...704A.298B}
}

@ARTICLE{Bellotti2024b,
       author = {{Bellotti}, S. and {Morin}, J. and {Lehmann}, L.~T. and {Petit}, P. and {Hussain}, G.~A.~J. and {Donati}, J. -F. and {Folsom}, C.~P. and {Carmona}, A. and {Martioli}, E. and {Klein}, B. and {Fouqu{\'e}}, P. and {Moutou}, C. and {Alencar}, S. and {Artigau}, E. and {Boisse}, I. and {Bouchy}, F. and {Bouvier}, J. and {Cook}, N.~J. and {Delfosse}, X. and {Doyon}, R. and {H{\'e}brard}, G.},
        title = "{Long-term monitoring of large-scale magnetic fields across optical and near-infrared domains with ESPaDOnS, Narval, and SPIRou. The cases of EV Lac, DS Leo, and CN Leo}",
      journal = {\aap},
         year = 2024,
        month = jun,
       volume = {686},
          eid = {A66},
        pages = {A66},
          doi = {10.1051/0004-6361/202348043},
archivePrefix = {arXiv},
       eprint = {2403.08590},
 primaryClass = {astro-ph.SR},
       adsurl = {https://ui.adsabs.harvard.edu/abs/2024A&A...686A..66B}
}

@book{KivelsonRussel,
    author = {Kivelson, M. G. and Russel, C. T.},
    title = {``Introduction to Space Physics"},
    publisher = {"Cambridge University Press"},
    year = "1995"
}

@article{Shields2016,
    author = {Shields, A. L. and Ballard, S and Johnson, J. A.},
    title = {``The habitability of planets orbiting M-dwarf stars"},
    journal = "Physics Reports",
    year = "2016",
    volume = "663",
    pages = "1--38"
}

\begin{appendix}
\nolinenumbers
\section{\swp{Accounting for unpolarized and linearly polarized light}}\label{appendix:StokesIVQU}
In the simulations performed in this work we have used ZDI maps obtained from circular polarizations ($V$)\swpt{.} For epoch 2023.7 of EV Lac, in addition to obtaining a ZDI map using circular polarization \cite{Donati2025} also obtained ZDI maps that used (1) both unpolarized and circularly polarized light ($IV$) and (2) unpolarized, circularly polarized, and linearly polarized light ($IVQU$). We performed simulations for these two maps using Solar value of $S_A/B$ and compared them against the simulation of this epoch of EV Lac that relied on circular polarization alone.

We summarized the characteristics of the stellar wind computed from these simulations in Table \ref{tab:evlac2023_comparison}. We found small differences in mass-loss rates between magnetic field maps obtained with the three approaches ($V$, $IV$, $IVQU$). The inclusion of unpolarized and linearly polarized light in the map decreases angular momentum-loss rate with respect to the circular polarization-only map by 1.6 times for the $IV$ map, and 1.3 for the $IVQU$ map. The Alfvén radii increase by a factor of 1.3-1.4 in simulations based on the $IV$ and $IVQU$ maps when compared to the simulation using the map reconstructed based on circular polarization alone. The increase in the Alfvén radius is expected since both the $IV$ and $IVQU$ maps have a stronger radial magnetic field ($\langle | B_r|\rangle$ of 405 G and 435 G respectively versus 281 G for the $V$ map).  The effect of including non-circular polarizations in the ZDI maps on simulation outputs should be explored more extensively once more stars get mapped with the technique used in \cite{Donati2025}.

\begin{table}[h!]
\centering
\caption{Comparison of simulation outputs for EV Lac 2023.7 ZDI map obtained with various polarizations.}
\resizebox{\linewidth}{!}{\begin{tabular}{lcccc}\hline\hline \\
Polarizations & \swp{$\dot{M}~[10^{-13}~{\rm M_\odot\,yr^{-1}}]$} & $\dot{J}~[10^{31}~{\rm erg}]$ & \swpt{$\langle R_A\rangle~[R_\star]$} & \swpt{$\langle \tilde{R_A}\rangle~[R_\star]$} \\ \hline
$V$    & 1.8 & 3.5 & 30 & 22 \\
$IV$   & 1.7 & 2.2 & 40 & 28 \\
$IVQU$ & 2.0 & 2.6 & 42 & 29 \\ \hline
\end{tabular}}
\label{tab:evlac2023_comparison}
\end{table}

\section{Weber-Davis model comparison}\label{appendix:poynting_WD}
\begin{figure}[ht!]
    \centering
    \includegraphics[width=1\hsize]{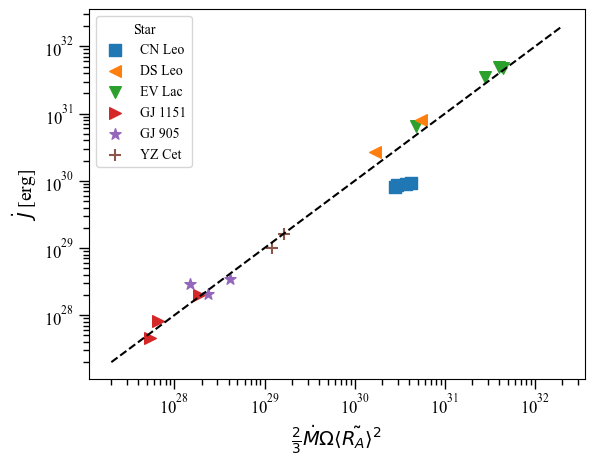}
    \caption{Comparison of our simulation outputs against the Weber-Davis model formula for angular momentum-loss rate. Multiple points per star indicate a different epoch. The dashed line is a 1-to-1 line}
    \label{fig:WeberDavis}
\end{figure}

We checked that our numerically computed angular momentum-loss rate (with Solar $S_A/B$) follows the Weber-Davis analytical model, which claims that $\dot{J} \propto \dot{M} \Omega \langle R_A\rangle^2$ where $\Omega$ is the angular speed of the star, and $\langle R_A\rangle$ is average Alfvén radius \citep[][]{WeberDavis1967}. It can be shown that \citep[for derivation see][]{Vidotto2021}:
\begin{equation}
    \dot{J} \approx \cfrac{2}{3} \dot{M} \Omega \langle \tilde{R_A}\rangle^2
\end{equation}
We carried out this calculation for all simulations. For $\dot{M}$ and $\langle \tilde{R_A}\rangle$, we use the numerically derived values as described in \swpt{Sect.} \ref{section:stellar_wind}. The comparison between the numerical angular momentum and the value computed with the model is shown in Fig. \ref{fig:WeberDavis}. With the exception of CN Leo, the estimates from our simulations are generally consistent with the values from the Weber-Davis model. The fundamental parameters of CN Leo (see Table \ref{tab:sample_properties}) are the most different from the Sun, hence it is not surprising to see a more significant deviation. We conclude this test noting that the simulations demonstrate expected behavior.

\end{appendix}

\end{document}